\documentclass{emulateapj}

\lefthead{van der Wel et al.}
\righthead{Comparing Dynamical and Photometric Masses at $z=1$}
\slugcomment{Accepted by the Astrophysical Journal}

\begin{document}

\title{Comparing Dynamical and Photometric Mass Estimates of Low- and 
High-Redsh ift Galaxies: Random and Systematic 
Uncertainties\altaffilmark{1,2,3}}

\author{A.~van~der~Wel\altaffilmark{4}, M.~Franx\altaffilmark{5},  S.~Wuyts\altaffilmark{5},
P.G.~van~Dokkum\altaffilmark{6},
J.~Huang\altaffilmark{7},
H.-W.~Rix\altaffilmark{8}, and
G.D. Illingworth\altaffilmark{9}}
\altaffiltext{1}{Based on observations collected at the European Southern 
Observatory, Chile (168.A-0485, 169.A-0458).}
\altaffiltext{2}{Based on observations with the 
\textit{Hubble Space Telescope}, obtained at the Space Telescope Science 
Institute, which is operated by AURA, Inc., under NASA contract NAS 5-26555}
\altaffiltext{3}{This work is based in part on observations made with the 
\textit{Spitzer Space Telescope}, which is operated by the Jet Propulsion 
Laboratory, California Institute of Technology under NASA contract 1407.}
\altaffiltext{4}{Department of Physics \& Astronomy, Johns Hopkins University, 3400 N. Charles Street, Baltimore, MD 21230}
\altaffiltext{5}{Leiden Observatory, P.O.Box 9513, NL-2300 AA, Leiden, 
The Netherlands}
\altaffiltext{6}{Department of Astronomy, Yale University, P.O. Box 208101, 
New Haven, CT 06520-8101}
\altaffiltext{7}{Harvard-Smithsonian Center for Astrophysics, 60 Garden Street,
Cambridge, MA 02138}
\altaffiltext{8}{Max-Planck-Institut f\"ur Astronomie, K\"onigstuhl 17, D-69117
Heidelberg, Germany}
\altaffiltext{9}{University of California Observatories/Lick Observatory, 
University of California, Santa Cruz, CA 95064}

\begin{abstract}

We determine the importance of redshift-dependent systematic
effects in the determination of stellar masses from broad
band spectral energy distributions (SEDs),
using high quality kinematic and photometric
data of early-type galaxies at $z\sim 1$ and $z\sim 0$.
We find that photometric masses of $z\sim 1$ galaxies can be
systematically different, by up to a factor of 2,
from photometric masses of $z\sim 0$ galaxies
with the same dynamical mass.
The magnitude of this bias depends on the choice of stellar
population synthesis model and the rest-frame wavelength
range used in the fits.
The best result, i.e., without significant bias,
is obtained when rest-frame optical
SEDs are fitted with  models from Bruzual \& Charlot (2003).
When the SEDs are extended to the rest-frame near-IR,
a bias is introduced:
photometric masses of the $z\sim 1$ galaxies increase
by a factor of 2 relative to the photometric masses
of the $z\sim 0$ galaxies.
When we use the Maraston (2005) models, the photometric masses
of the $z\sim 1$ galaxies are low relative
to the photometric masses of the $z\sim 0$ galaxies by a factor of 
$\sim 1.8$.
This offset occurs both for fits based on rest-frame optical SEDs, and fits
based on rest-frame optical+near-IR SEDs.
The results indicate that model uncertainties produce uncertainties as high
as a factor of 2.5
in mass estimates from rest-frame near-IR photometry, independent of
uncertainties due to unknown star formation histories.
\end{abstract}


\section{Introduction}\label{intro}
Galaxy masses are an essential link between theories for galaxy formation 
and observations of the galaxy population and the evolution thereof.
In the local universe masses can be measured accurately by 
modeling the luminosity distribution and the dynamical structure of galaxies
\citep[see, e.g.,][]{cappellari05}.
Scaling relations such as the fundamental plane for early-type galaxies
\citep{djorgovskidavis87,dressler87}
and the Tully-Fisher relation for spiral galaxies
\citep{tully77}
can be used to measure the evolution of the mass-to-light ratio ($M/L$)
from high redshift to the present
\citep{franx93}.
This technique has been applied successfully and has provided
constraints on the formation epoch of massive early-type galaxies out to $z\sim 1.3$
\citep[e.g.,][]{vandokkumstanford03}.

The number of galaxies with dynamically measured masses 
at intermediate redshifts ($z\sim 1$) is small, because obtaining those is 
observationally expensive. Furthermore, those samples are severely
hampered by selection effects
\citep[see][]{vanderwel05}. Therefore, it is not yet possible
to obtain directly the galaxy mass density at high redshift.
Also, even though the redshift at which dynamical masses can be measured
is steadily increasing, the most active era of galaxy formation, $z\ge 2$,
is not accessible in that respect with the current generation of instruments. 
For all of these reasons, one has to rely on less accurate mass estimates to 
construct a picture of the high-$z$ galaxy population
\citep{shapley03,papovich03,forster04,labbe05,shapley05}
and the evolution of the mass density with redshift
\citep{bell04b,drory04,faber05}. 
In these studies, broadband photometry is compared with 
predictions from stellar population models
\citep[most commonly,][]{bruzual03}
in order to constrain the physical properties of high $z$-galaxies, and 
thereby their stellar masses.

However, the uncertainties in mass estimates obtained by fitting the spectral energy distributions
(SEDs) of galaxies are large. It is well-known that 
parameters such as age, dust content and metallicity are degenerate,
leaving $M/L$ uncertain. The lack of knowledge of the stellar initial mass function (IMF)
leads to a (probably large) systematic uncertainty.
Comparisons of photometric mass estimates with dynamical mass measurements
are essential to establish their robustness and accuracy.
\citet{bell01} have shown that the optical colors of spiral galaxies
correlate well with their $M/L$. 
At $z\sim 1$ \citet{vanderwel05} have demonstrated that optical
colors correlate well with dynamically determined $M/L$ for early-type galaxies.
These results suggest that full SEDs should provide good constraints 
on masses and $M/L$.
\begin{deluxetable*}{lccccccccccc}
\tabletypesize{\scriptsize}
\tablecolumns{11}
\tablewidth{0pt}
\tablenum{1}
\tablecaption{Photometry}
\tablehead {
\colhead{ID} &
\colhead{$b_{435}$} &
\colhead{$v_{606}$} &
\colhead{$i_{775}$} &
\colhead{$z_{850}$} &
\colhead{$J$} &
\colhead{$K$} &
\colhead{$3.6\mu$} &
\colhead{$4.5\mu$} &
\colhead{$5.8\mu$} &
\colhead{$8.0\mu$}\\
\colhead{} &
\colhead{} &
\colhead{} &
\colhead{} &
\colhead{} &
\colhead{} &
\colhead{} &
\colhead{} &
\colhead{} &
\colhead{} &
\colhead{}}
\startdata
 CDFS-1 & $24.49 \pm 0.12$ & $23.79 \pm 0.11$ & $22.10$ & $20.95$ & $19.95$ & $18.12$ & $16.68$ & $16.46$ & $16.44 \pm 0.08$ & $15.73 \pm 0.08$  \\
 CDFS-2 & $24.95 \pm 0.18$ & $23.33 \pm 0.07$ & $21.51$ & $20.45$ & $19.52$ & $17.79$ & $16.54$ & $16.40$ & $16.43 \pm 0.08$ & $15.87 \pm 0.09$  \\
 CDFS-3 & $25.70 \pm 0.33$ & $23.69 \pm 0.10$ & $22.47$ & $21.44$ & $20.53$ & $18.84$ & $17.47$ & $17.05$ & $17.50 \pm 0.21$ & $17.03 \pm 0.24$  \\
 CDFS-4 & $24.02 \pm 0.08$ & $22.70 \pm 0.04$ & $21.01$ & $19.98$ & $18.88$ & $17.04$ & $15.79$ & $15.62$ & $15.53 \pm 0.04$ & $15.27 \pm 0.05$  \\
 CDFS-5 & $>26.4$          & $23.90 \pm 0.12$ & $21.91$ & $21.24$ & $20.24$ & $18.52$ & $17.53$ & $17.60$ & $17.45 \pm 0.20$ & $17.98 \pm 0.50$  \\
 CDFS-6 & $25.23 \pm 0.23$ & $22.66 \pm 0.04$ & $20.98$ & $20.39$ & $19.57$ & $17.94$ & $17.18$ & $17.08$ & $16.96 \pm 0.13$ & $16.87 \pm 0.21$  \\
 CDFS-7 & $25.32 \pm 0.25$ & $23.48 \pm 0.08$ & $22.01$ & $20.92$ & $19.78$ & $17.86$ & $16.39$ & $16.14$ & $15.83 \pm 0.05$ & $15.95 \pm 0.09$  \\
CDFS-12 & $24.41 \pm 0.11$ & $23.75 \pm 0.10$ & $22.24$ & $21.28$ & $20.31$ & $18.50$ & $17.15$ & $16.93$ & $17.04 \pm 0.14$ & $16.45 \pm 0.15$  \\
CDFS-13 & $25.49 \pm 0.28$ & $23.14 \pm 0.06$ & $21.40$ & $20.42$ & $19.41$ & $17.62$ & $16.50$ & $16.36$ & $16.39 \pm 0.08$ & $16.26 \pm 0.12$  \\
CDFS-14 & $24.97 \pm 0.18$ & $23.28 \pm 0.07$ & $21.56$ & $20.55$ & $19.56$ & $17.79$ & $16.37$ & $16.33$ & $16.15 \pm 0.06$ & $15.97 \pm 0.10$  \\
CDFS-15 & $24.89 \pm 0.17$ & $22.67 \pm 0.04$ & $20.99$ & $20.37$ & $19.51$ & $17.85$ & $16.99$ & $17.08$ & $16.45 \pm 0.08$ & $16.83 \pm 0.20$  \\
CDFS-16 & $25.07 \pm 0.20$ & $22.67 \pm 0.04$ & $21.02$ & $20.34$ & $19.50$ & $17.73$ & $16.81$ & $16.87$ & $16.87 \pm 0.12$ & $16.87 \pm 0.21$  \\
CDFS-18 & $24.60 \pm 0.13$ & $23.03 \pm 0.05$ & $21.48$ & $20.46$ & $19.36$ & $17.52$ & $16.19$ & $16.03$ & $15.95 \pm 0.05$ & $15.66 \pm 0.07$  \\
CDFS-19 & $23.52 \pm 0.05$ & $22.17 \pm 0.02$ & $20.86$ & $20.17$ & $19.32$ & $17.62$ & $16.43$ & $16.23$ & $16.01 \pm 0.06$ & $15.84 \pm 0.08$  \\
CDFS-20 & $26.20 \pm 0.49$ & $23.34 \pm 0.07$ & $21.62$ & $20.46$ & $19.42$ & $17.50$ & $16.14$ & $15.96$ & $16.07 \pm 0.06$ & $15.60 \pm 0.07$  \\
CDFS-21 & $24.30 \pm 0.10$ & $22.86 \pm 0.05$ & $21.18$ & $20.58$ & $19.78$ & $18.19$ & $17.05$ & $17.02$ & $16.67 \pm 0.10$ & $16.36 \pm 0.13$  \\
CDFS-22 & $25.23 \pm 0.23$ & $22.68 \pm 0.04$ & $20.81$ & $20.08$ & $19.12$ & $17.38$ & $16.33$ & $16.41$ & $16.22 \pm 0.07$ & $16.10 \pm 0.11$  \\
CDFS-23 & $26.12 \pm 0.46$ & $25.43 \pm 0.41$ & $23.12$ & $21.91$ & $20.84$ & $19.16$ & $17.64$ & $17.46$ & $17.54 \pm 0.22$ & $16.67 \pm 0.18$  \\
CDFS-25 & $>26.4$          & $24.49 \pm 0.19$ & $22.58$ & $21.42$ & $20.34$ & $18.76$ & $17.37$ & $17.28$ & $16.89 \pm 0.13$ & $16.66 \pm 0.17$  \\
CDFS-29 & \nodata          & $22.43 \pm 0.03$ & $21.40$ & $20.46$ & $19.65$ & $18.10$ & $16.79$ & $16.53$ & $16.75 \pm 0.11$ & $15.86 \pm 0.09$  \\
\enddata

\tablecomments{
Photometry of the high-redshift field galaxy sample.
IDs are the same as in van der Wel et al. (2005b).
All magnitudes within a $5''$-diameter aperture of PSF-matched images.
The two lower-limits that occur in the table are $3\sigma$-limits.
Object CDFS-29 falls outside the $b_{435}$ ACS mosaic.
The photometry in the columns without listed errors have errors 
on all individual objects of 0.05 mag or less.
The typical errors on the $i_{775}$, $z_{850}$, $J$, $K$, $3.6\mu$, and $4.5\mu$
data points are 0.02, 0.01, 0.01, 0.01, 0.01, and 0.01 mag, respectively.
\label{tab1}}
\end{deluxetable*}
This has been tested directly for local galaxies by \citet{drory04b},
who establish the reliability of the ``photometric masses''.

In this paper we extend, for the first time, the comparison between 
dynamical and photometric masses to $z=1$.
We use our sample of early-types with velocity dispersions \citep{vanderwel05},
and rest-frame UV-through-IR photometry.
We investigate the scatter between dynamical and photometric masses, 
both at $z=0$ and $z=1$, and how this depends on the photometry included
in the SED fits, and the model assumptions.
Furthermore, we investigate systematic differences between dynamical and 
photometric masses.
We note that photometric masses are never absolute, as dark 
matter fractions and numbers low-mass stars are unconstrained from the
photometry \citep[see, e.g.,][]{gerhard01,cappellari05}. 
However, photometric masses have been used to obtain
"relative masses" at higher redshifts, and to determine the relative
mass evolution 
\citep[e.g.,][]{dickinson03,rudnick03,bell04b,rudnick06}.
We test here explicitly whether the relative photometric masses 
at $z=0$ and $z=1$ are consistent with the dynamical masses.
The only assumption we make is that 
high- and low-redshift early-types have the same kind of stellar population
(IMF, metallicity), with  (obviously) different ages.
This assumption underlies all work with relative photometric masses.

We specifically address the question whether the rest-frame near-infrared 
(near-IR) helps to constrain the masses.
Because the near-IR is less sensitive to extinction than the optical,
extending SED fitting to the near-IR helps to lift the degeneracy between 
age and extinction.
On the other hand, the stellar population models are less reliable at 
wavelengths longer than $1\mu$ than in the optical \citep{maraston05}.
Furthermore, optical-to-near-IR colors have been shown to 
correlate less well with $M/L$ than optical colors \citep{bell01}.
The advance of IRAC \citep{fazio04} provides access to the rest-frame IR
at high redshift, and it is assumed that this will allow for 'cheap' mass determinations
through SED fitting for large numbers of distant galaxies.
It is clear that extending the fit SEDs to the IR and thus obtaining
mass estimates needs to be tested.

This paper is organized as follows:
we describe the dynamical masses of the galaxy samples in Section \ref{dyn}.
The photometry, the stellar population models, and our fitting method to 
obtain stellar masses are described in Section \ref{phot}.
In Section \ref{results} we present our results, discussing the
consistency of the models with our empirical results, and the dependence on the
fit wavelength range.
In Section \ref{discussion} we discuss the biases that are revealed 
by this work, and how this affects estimates of
high-$z$ galaxy masses and the evolution of the mass density.
Throughout we use the Vega magnitude system
(based on the \citet{kurucz92} A0V model spectrum), 
and the concordance cosmology,
$(\Omega_M,\Omega_{\Lambda},h)=(0.3,0.7,0.7)$.

\section{Dynamical Masses of Early-Type Galaxies}\label{dyn}
Dynamical masses are computed as
$M_{dyn}=Cr_{\rm{eff}}\sigma_c^2 /G$,
where $C$ is a constant,
$\sigma_c$ is the central velocity dispersion, 
$r_{\rm{eff}}$ is the effective radius and
$G$ is the gravitational constant.
\citet{kochanek94},
assuming spherical symmetry, no rotation and fixed anisotropy,
has shown that the velocity dispersion of the
dark matter in elliptical galaxies equals the central line-of-sight 
velocity dispersion, as measured within a $2''\times 4''$ sized aperture.
This implies $C=4$ if $M_{dyn}$ is 
assumed to be twice the mass within the effective radius.
In terms of the dispersions we use
(corrected to an $3\farcs4$-diameter aperture at the distance
of the Coma cluster), we would have to use $C=4.11$, taking
the differences in aperture and distance between the samples
into account.
\begin{figure*}[t]
\begin{center}
\leavevmode
\hbox{%
\epsfxsize=18cm
\epsffile{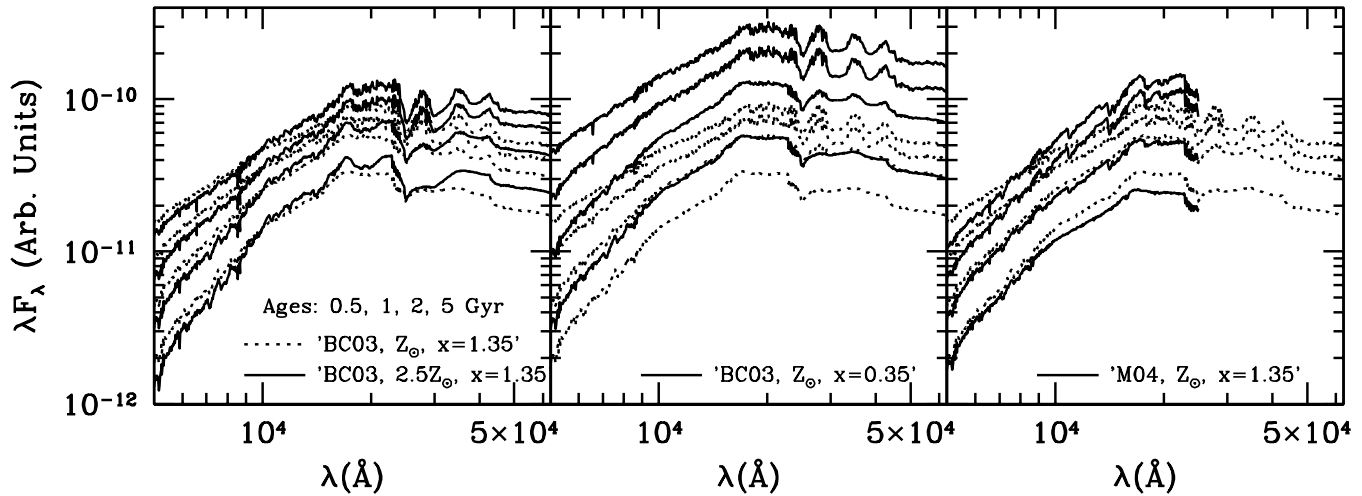}}
\figcaption{\small
Various model spectra for a range of ages and parameters.
The labels 'BC03' identify Bruzual \& Charlot (2003) models; 
the label 'M05' identifies the Maraston (2005) model.
Metallicity and IMF slope are also indicated ($x=1.35$ corresponds
to the Salpeter IMF).
The indicated ages run from top to bottom.
M05 does not provide realistic spectra for $\lambda>2.5\mu$.
Note the difference in optical-to-near-IR slope
between the BC03 and M05 models.
\small
\label{models}}
\end{center}
\end{figure*}
In the literature, $C=5$ is used more often
\citep[e.g.,][]{jorgensen96}.
Therefore, we choose to use $C=5$ throughout this paper in order for the 
masses to remain consistent with previous studies.
This number is based on an isotropic density model,
projected onto an $r^{1/4}$-law and integrated out to $2.5r_{eff}$,
which 75\% of the total mass.
\citep[see, e.g.,][]{cappellari05}.
In units of solar masses, the dynamical mass becomes
$M/M_{\odot}=1.17\times10^6~r_{\rm{eff}} \sigma_c^2$,
with $r_{\rm{eff}}$ in $kpc$ and $\sigma_c$ in $km~s^{-1}$.
We note, however, that the precise value of $C$ does
not affect our analysis as long as it does not evolve with redshift.

\citet{vanderwel05} provide internal velocity dispersions
and structural parameters for 29 galaxies in the Chandra Deep Field-South
(CDF-S).
We exclude four galaxies with late-type morphologies, four
early-types with spectra with low signal-to-noise ratios ($S/N<12~\AA^{-1}$),
and one early-type that falls outside the IRAC mosaic.  We then have a sample
of 20 early-type galaxies with a median redshift of $z=0.98$.
The median mass of this sample is $1.5\times 10^{11}M_{\odot}$.
As a low-redshift comparison sample we use 23 
early-type galaxies in the nearby ($z=0.024$)
Coma cluster with measured $\sigma_c$ and $r_{\rm{eff}}$
\citep{dressler87,faber89}.
The median mass of this sample, $M=1.8\times10^{11}M_{\odot}$,  
is comparable to that of the $z\sim 1$ sample.

\section{Derivation of Stellar Masses from Photometry}\label{phot}

\subsection{Photometric Data}
A large range of photometric data is available for the CDF-S.
GOODS provides ACS imaging in 4 filters (F435W, F606W, F775W, F850LP,
hereafter $b_{435}$, $v_{606}$, $i_{775}$, $z_{850}$
\citep{giavalisco04}), ESO provides
$J$- and $K$-band imaging, and IRAC GTO observations from the 
\textit{Spitzer Space Telescope} are available 
(channels 1-4; $3.6\mu$, $4.5\mu$, $5.8\mu$, and $8.0\mu$, respectively).
The photometry is described by \citet{vanderwel06} 
and is presented here in Table \ref{tab1}.
The magnitudes are measured within a fixed aperture 
with a diameter of $5\farcs0$, using
registered and PSF-matched images.

For the local sample of Coma galaxies we need photometry
that samples a similar rest-frame wavelength range as 
we have for the $z\sim 1$ sample.
\citet{faber89} and \citet{scodeggio98} provide the optical surface photometry,
and \citet{pahre98} provide the $K$-band surface photometry.
We use the effective surface brightnesses and effective radii in
the $B$-, $V$-, $I$-, and $K$-bands to compute the colors
within apertures of 67'', used by Faber et al., which typically corresponds
to $4r_{\rm{eff}}$ at the distance of the Coma cluster. 
The average colors are 
$B-V=0.96$, $V-I=1.18$, and $I-K=1.96$.
The Coma galaxies are the only sample in the literature with K-band 
photometry suitable for this study. The fact that these are cluster
galaxies, and not field galaxies as the galaxies in our distant sample 
are, does not limit the interpretation of our results.
The differences between the two samples are reproduced through modeling
their SEDs, adopting the relevant parameters which might differ (such as 
age, SFH, and dust content) as free parameters in the fit.
The only assumption is that they have similar stellar 
populations in terms of IMF and metallicity.

\subsection{Stellar Population Models}\label{secmodels}
For our purpose we need stellar population models that provide
synthetic spectra over a large wavelength range ($0.3-4\mu$).
In recent years,
\citet{bruzual03}
(hereafter, BC03)
and
\citet{maraston05}
(hereafter, M05),
have provided such spectra.
M05 provide realistic spectra only up to $2.5\mu$, 
limiting our SED fitting range to the $4.5\mu$-channel. This
is not a severe limitation as the photometric
errors in the two longest-wavelength channels are much larger than in the other two channels
(see Table \ref{tab1}), and observed $4.5\mu$ at $z\sim 1$
corresponds to the reddest filter ($K$) available
for the nearby sample.

Significant differences between the models occur for all ages of interest
for this study. 
For ages $0.5-2~\rm{Gyr}$, the M05 model is much redder in the optical-to-nearIR 
than BC03, because of the very different implementation of the 
Thermally-Pulsating Asymptotic Giant Branch (TP-AGB) phase.
\null
\vbox{
\begin{center}
\leavevmode
\hbox{%
\epsfysize=9cm
\epsffile{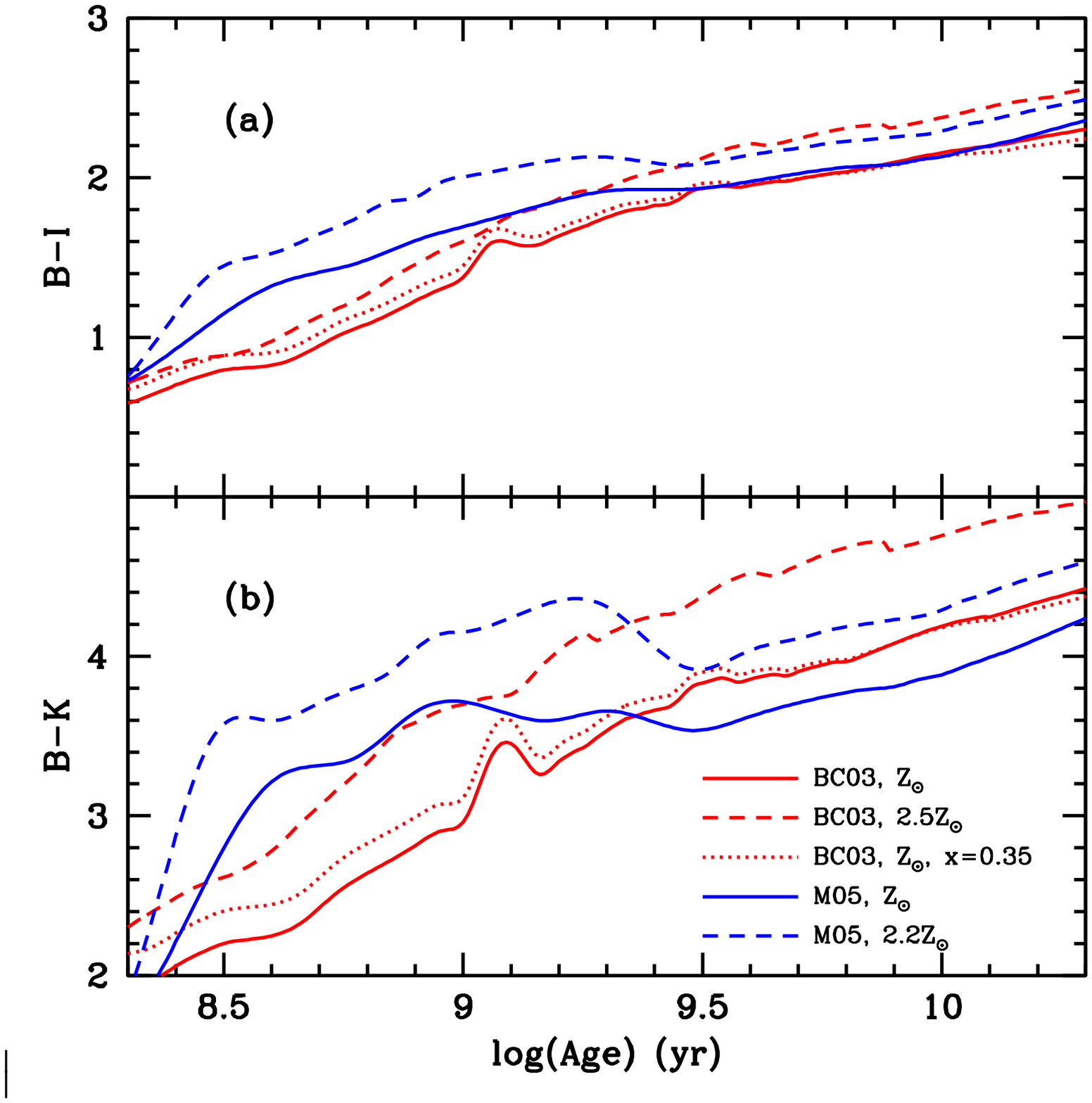}}
\figcaption{\small
Evolution of $B-I$ (a) and $B-K$ (b) with age 
for different models and different model parameters.
The differences between the BC03 and M05 models for similar 
sets of parameters are striking.
For ages $<3$ Gyr differences up to 0.5 mag occur in $B-I$, 
in $B-K$ even up to 1.4 mag.
For older ages, the agreement is good in $B-I$,
whereas in $B-K$ the difference is $\sim 0.5$ mag.
The tracks are limited to ages from 0.5-20 Gyr.
\small
\label{age_col}}
\end{center}}
For ages older than ~3 Gyr, the M05 model is bluer 
in optical-to-near-IR colors than the BC03 model because of the cooler 
Red Giant Branch (RGB) stars of the Padova tracks \citep[][used by BC03]{girardi00},
with respect to those of the Cassisi tracks \citep[][used by M05]{cassisi00}.
M05 adopt
the fuel consumption approach, and calculates luminosity contributions
from different stellar types by the amount of fuel used during a
certain evolutionary stage,
whereas BC03 follow the isochrone synthesis approach, and characterize
the properties of the stellar population per mass bin.

We consider SSP models from M05 with solar and super-solar metallicity 
($Z_{\odot}$ and $2.2Z_{\odot}$) and a Salpeter IMF
with mass limits $0.1M_{\odot}$ and $100M_{\odot}$ 
(these limits are used throughout the paper).
For the BC03 model we also consider models with two different metallicities
($Z_{\odot}$ and $2.5Z_{\odot}$) and, in addition,
a solar-metallicity model with a 'top-heavy', or 'flat' IMF with slope of 
$x=0.35$ (instead of $x=1.35$, which is the Salpeter IMF).
The model with a flat IMF has been 
shown to provide a better match to the evolution of $M/L$ in the optical
and in the near-IR simultaneously
\citep{vanderwel06}.
We do not consider IMFs with different shapes at the low-mass end. 
The reason is that varying this shape  changes all mass estimates by a constant factor. In other words, the high- and low-redshift
samples are affected similarly by the choice of parametrization
of the low-mass end of the IMF.

In Figure \ref{models} we show the spectra of the three BC03 models and the 
M05 model for ages of 0.5, 1, 2, and 5 $Gyr$.
The difference between the BC03 models with different metallicities,
shown in the left-hand panel, is subtle,
except for the longest wavelengths.
The BC03 models with different IMF slopes (middle panel) show much larger 
variation. This is primarily a difference in overall energy output, 
which is simply due to the larger numbers of giants in the case of a flat IMF, 
especially at young ages.
A secondary, but actually more interesting difference is the different 
evolution of the slope, i.e., color, of the SED in models with different IMFs. 
Such differences in color evolution are also apparent when comparing the 
BC03 models to those of M05, shown in the right-hand panel.
For ages younger than $2.5~Gyr$, the M05 model predicts much
higher luminosities in the $K$-band than the BC03 model, whereas the optical
luminosities are similar. 
The differences in color evolution are demonstrated more clearly in 
Figure \ref{age_col}.
For ages older than 3 Gyr the optical colors from different models agree well.
For ages younger than that differences of up to 1 magnitude in $B-I$ 
occur between BC03 and M05 models with the same model parameters.
For the near-IR differences occur for all ages and can increase up to
1.4 magnitudes in $B-K$.

\subsection{Fitting Method}\label{method}
We use redshifted model spectra to compute apparent magnitudes, $m_{mod}$,
allowing the age (and, optionally, the star-formation history (SFH) and the 
dust content) to vary, and normalizing the calculated magnitudes
to match the observed magnitudes, $m_{obs}$, to obtain the photometric mass.
$m_{obs}$ are the color magnitudes given in Table \ref{tab1},
including a correction from fixed aperture magnitudes to total
magnitudes as measured in the $K$-band.
The resulting total fluxes are multiplied by 0.75
to match the aperture in which the dynamical mass is calculated 
(see Section \ref{dyn}).
The derived stellar mass includes the dark, compact remnants of 
massive stars, but not the gas lost due to stellar winds and supernovae ejecta.
The best-fitting model is selected on the basis of the root mean square 
of $m_{mod}-m_{obs}$ ($\rm{RMS}$), weighing with the inverse square
of the photometric errors. A certain minimum error is assumed to avoid that the
data points with the largest errors are effectively ignored in the SED fits.
When the IRAC data are used in the fit, a minimum error of 0.10 mag is 
adopted. When the IRAC data are not used, we use 0.03 mag. The choices are 
based on the relative uncertainties in the photometric
zeropoints of the different datasets.
However, the precise value 
of the minimum error does not profoundly affect the fitting results
for the samples as a whole

For the SSP models described in the previous section,
there is a unique relation between color and age. Subsequently, age
determines $M/L$, and thus $M_{phot}$.
The differences between the models, discussed in the previous section
and shown in Figure \ref{age_col},
will cause mass estimates derived from SED to differ.
Most notably, the different trends with age will cause
systematic biases in mass estimates of young (high-$z$) galaxies
relative to old (low-$z$) galaxies.

In Figure \ref{rms} we show, as an example, the fitting results for object CDFS-2,
using the SSP models.
In the top panel we show how $\log(M_{phot})$ of the best-fitting BC03 model 
(with solar metallicity and a Salpeter IMF)
changes if different wavelength ranges are used in the fit.
\null
\vbox{
\begin{center}
\leavevmode
\hbox{%
\epsfysize=9cm
\epsffile{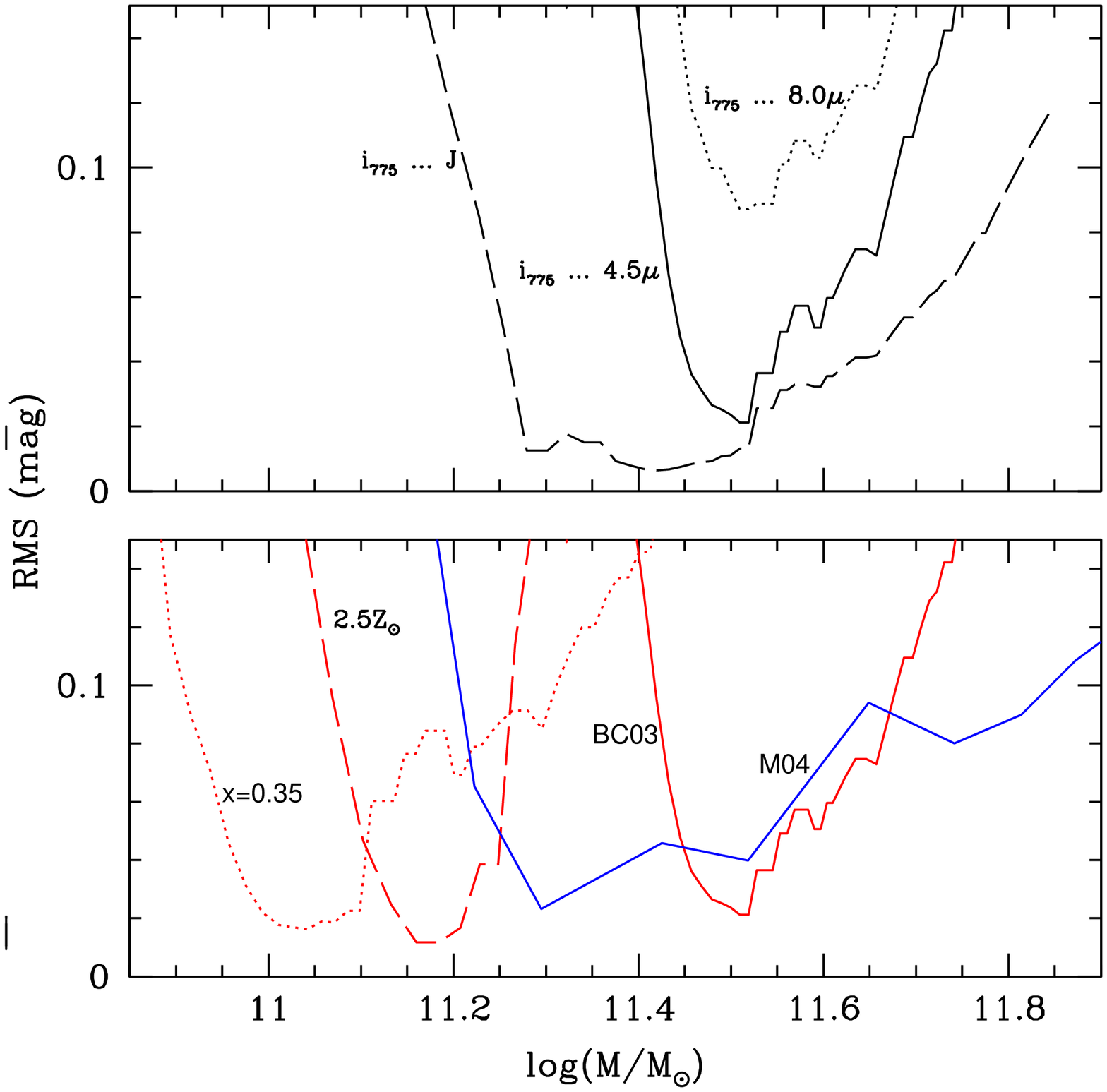}}
\figcaption{\small
The quality of the SED fit, quantified by the root mean square ($\rm{RMS}$) of
the difference between the calculated model magnitudes and the observed magnitudes,
of object CDFS-2 at $z=0.96$, as an illustration of the fitting method.
The upper panel shows how the results vary by changing the
wavelength range included in the fit, using the BC03 model
with solar metallicity and a Salpeter IMF.
\textit{dashed line:} result for fitting $i_{775}$, $z_{850}$, and $J$.
\textit{solid line:} result for fitting $i_{775}$, $z_{850}$, $J$, $K$, $3.6\mu$ and $4.5\mu$.
\textit{dotted line:} result for fitting $i_{775}$, $z_{850}$, $J$, $K$, $3.6\mu$, $4.5\mu$, $5.8\mu$ and $8.0\mu$.
Including the rest-frame near-IR in the fit leads to a somewhat higher $M/L$.
The bottom panel shows how the results vary when modeling $i_{775}$ through $4.5\mu$
using different models.
\textit{solid red line:} result for the BC03 model with Solar metallicity and a Salpeter IMF.
\textit{dashed red line:} result for the BC03 model with super-solar metallicity.
\textit{dotted red line:} result for the BC03 model with a top-heavy IMF.
\textit{solid blue line:} result for the M05 model with solar metallicity and a Salpeter IMF.
Different models clearly yield different $M/L$ ratios.
The broad minimum obtained for the M05 model can be explained by the slow
evolution of the optical-to-near-IR color (see Figure \ref{age_col}), 
and the consequently poorly constrained age.
\small
\label{rms}}
\end{center}}
In the bottom panel we show how the results vary from model to model,
fitting the SEDs from $i_{775}$ to $4.5\mu$.
$\log(M_{phot})$ of the best-fitting models vary by a factor of 3,
indicating the level of the systematic uncertainty in the photometric
mass estimate.
Note that the quality of the fits is generally good: 
the $\rm{RMS}$ of the best-fitting model is typically only $\sim 0.02$ mag. 

To further illustrate our SED fitting method, 
we show in Figure \ref{seds} 
the SEDs of three $z\sim 1$ early-types and two different, best fitting model spectra
for each of those. Those model spectra are the SSP models with solar 
metallicity and a Salpeter IMF of BC03 and M05, which are fit 
to the $i_{775}$, $z_{850}$, and $J$ data-points, i.e., the rest-frame optical.
While similar in the rest-frame optical, 
the two models differ from each other in the rest-frame near-infrared.
\null
\vbox{
\begin{center}
\leavevmode
\hbox{%
\epsfysize=9cm
\epsffile{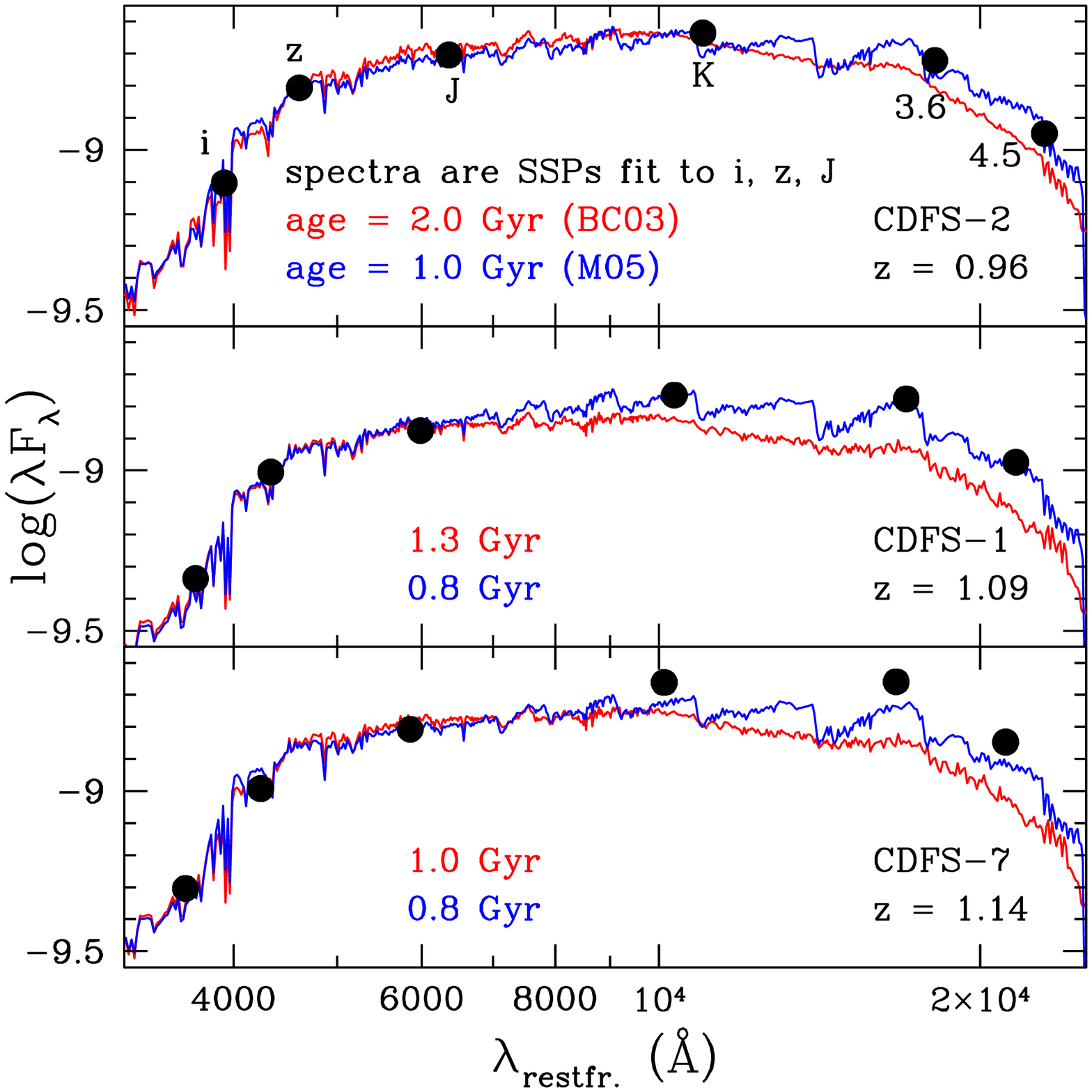}}
\figcaption{\small
SEDs and model spectra of three typical $z\sim 1$ early-type galaxies.
The model spectra, which are for SSPs
with solar metallicity and a Salpeter IMF,
are fit to the three shortest wavelength data-points shown here
($i_{775}$, $z_{850}$, and $J$), with age as the only free parameter besides
the normalization.
The BC03 model under-predicts the fluxes in the redder bands, whereas
the M05 model makes better predictions.
\small
\label{seds}}
\end{center}}
The BC03 spectrum predicts systematically lower flux levels than the M05 model. 
The $K$, $3.6\mu m$, and $4.5\mu m$ data-points agree better with the M05 model
than with the BC03 model.
As a consequence, when we fit the $i_{775}$ through $4.5\mu m$ SEDs of these galaxies,
the best-fitting age increases for the BC03 model in order to match
the redness of the SEDs. No older ages are found when the M05 is used.
This behavior is typical for the galaxies in our sample.
As we will show in the subsequent sections, whether or nor including
the near-infrared in the SED fits has profound consequences on the
resulting photometric mass estimates.

\section{Systematic Uncertainties in Photometric Mass Estimates}\label{results}
\subsection{Rest-Frame Optical SED Fits}\label{opt}
In this section we compare photometric and dynamical mass estimates.
We begin by fitting the rest-frame optical SEDs, as previous work
has shown that a single rest-frame optical color like $B-R$ can
be used well to estimate $M/L$ \citep{bell01}
For the local sample we use the $B$-, $V$-, and $I$-bands in the fit,
and for the distant sample we use $i_{775}$, $z_{850}$, and $J$.

\begin{figure*}[t]
\begin{center}
\leavevmode
\hbox{%
\epsfxsize=15cm
\epsffile{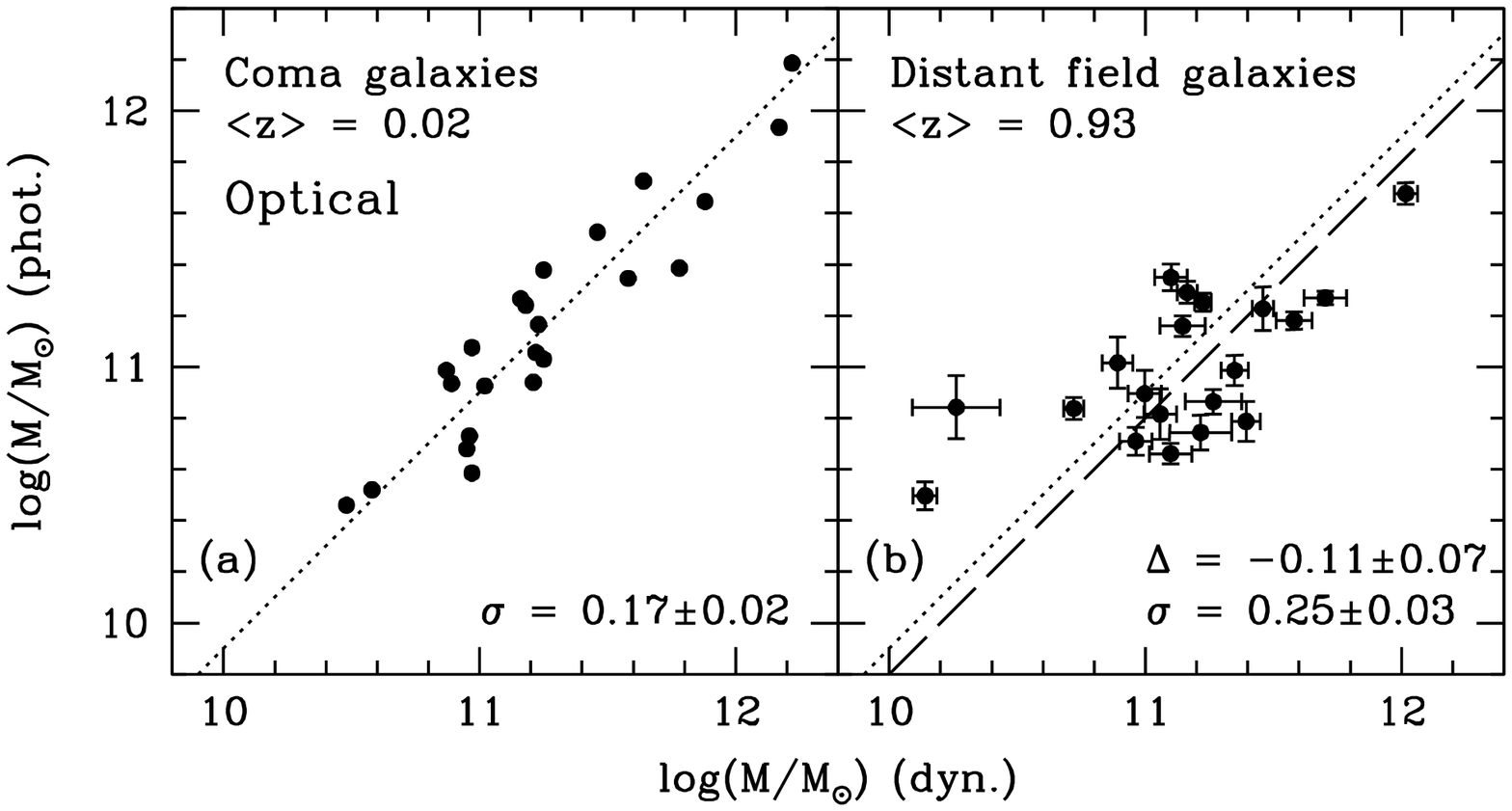}}
\figcaption{\small
(a) Comparison between dynamical and photometric masses
of early-type galaxies in the Coma cluster at $z=0.024$.
The photometric masses are obtained by fitting the photometric SEDs
($B$, $V$, $I$) by model spectra from BC03 for 
a SSP with solar metallicity and a Salpeter IMF.
Age and mass are the only free parameters.
The dotted line indicates the best-fitting line with slope 1,
indicating the average deviation. 
$\sigma$ is the standard deviation from that line.
(b) Comparison between dynamical and photometric masses
of the distant field early-type galaxies. 
Photometric masses are obtained by fitting the photometric SEDs
from $i_{775}$, $z_{850}$, and $J$ by the same BC03 model as used for the local galaxies.
The dotted line, the same as in panel (a), is shown as a reference.
The dashed line is the best-fitting line with slope 1 to the distant
field sample, excluding two galaxies
with velocity dispersions lower than $100~km~s^{-1}$.
The errors on the photometric masses are obtained through Monte-Carlo 
simulations. 
\small
\label{MMstandardopt}}
\end{center}
\end{figure*}

We start with mass estimates resulting from fits  of  the SSP
model from BC03 with solar metallicity and a Salpeter IMF.
The left-hand panel of Figure \ref{MMstandardopt} shows
$M_{dyn}$ vs. $M_{phot}$ for the sample of local early-type galaxies.
On the right-hand side we show the same relation for the distant galaxies.

\begin{deluxetable*}{lccccccc}[b]
\tabletypesize{\scriptsize}
\tablecolumns{8}
\tablewidth{0pt}
\tablenum{2}
\tablecaption{SED fitting results}
\tablehead {
\colhead{Model} &
\colhead{$\Delta$} &
\colhead{$\sigma_1$} &
\colhead{$\delta_1$} &
\colhead{$<\rm{rms}>$} &
\colhead{$<Age>$}&
\colhead{$<SFH>$} &
\colhead{$<A_V>$}\\
\colhead{} &
\colhead{} &
\colhead{} &
\colhead{} &
\colhead{$mag$} &
\colhead{$Gyr$} &
\colhead{$Gyr$ or \%} &
\colhead{$mag$}}
\startdata
\multicolumn{8}{c}{$z=1$: $i_{775}$ , $z_{850}$ , $J$}\\
\hline
BC03, $Z_{\odot}$           &  $ -0.11\pm 0.07 $ & $ 0.25 \pm 0.03  $ & $ -0.20 \pm 0.07 $ &    0.03 & 2.0 & \nodata & \nodata \\
BC03, $Z_{\odot}$, $x=0.35$ &  $ -0.37\pm 0.07 $ & $ 0.27 \pm 0.03  $ & $ -0.70 \pm 0.06 $ &    0.03 & 1.8 & \nodata & \nodata \\
BC03, $2.5Z_{\odot}$        &  $  0.00\pm 0.06 $ & $ 0.24 \pm 0.02  $ & $ -0.32 \pm 0.05 $ &    0.04 & 1.1 & \nodata & \nodata \\
M05, $Z_{\odot}$            &  $ -0.23 \pm 0.06 $ & $ 0.22 \pm 0.03  $ & $ -0.33 \pm 0.05 $ &   0.05 & 1.7 & \nodata & \nodata \\
M05, $2.2Z_{\odot}$         &  $ -0.08 \pm 0.08 $ & $ 0.27 \pm 0.03  $ & $ -0.48 \pm 0.06 $ &   0.05 & 0.7 & \nodata & \nodata \\
\hline
\multicolumn{8}{c}{$z=1$: $i_{775}$ , $z_{850}$ , $J$ , $K$ , $3.6\mu$ , $4.5\mu$}\\
\hline
BC03, $Z_{\odot}$                          &  $ 0.15 \pm 0.06 $ & $ 0.22 \pm 0.04  $ & $ 0.08  \pm 0.05 $ &  0.10 & 3.9 & \nodata & \nodata \\
BC03, $Z_{\odot}$, $\tau$                  &  $ 0.03 \pm 0.06 $ & $ 0.23 \pm 0.03  $ & $ -0.04 \pm 0.03 $ &  0.04 & 3.0 & 0.97    & 0.56    \\
BC03, $Z_{\odot}$, double burst (10\%)     &  $ 0.07 \pm 0.06 $ & $ 0.23 \pm 0.04  $ & $  0.00 \pm 0.05 $ &  0.07 & 2.9 & 10\%    & 3.52$^*$  \\
BC03, $Z_{\odot}$, double burst (30\%)     &  $ 0.01 \pm 0.07 $ & $ 0.24 \pm 0.03  $ & $ -0.06 \pm 0.04 $ &  0.06 & 2.5 & 30\%    & 1.87$^*$  \\
BC03, $Z_{\odot}$, $x=0.35$                &  $ -0.02\pm 0.07 $ & $ 0.22 \pm 0.03  $ & $ -0.39 \pm 0.05 $ &  0.09 & 3.4 & \nodata & \nodata \\
BC03, $2.5Z_{\odot}$                       &  $ 0.23 \pm 0.07 $ & $ 0.25 \pm 0.04  $ & $ -0.23 \pm 0.06 $ &  0.08 & 1.3 & \nodata & \nodata \\
M05, $Z_{\odot}$                           &  $ -0.26 \pm 0.07$ & $ 0.24 \pm 0.04  $ & $ -0.19 \pm 0.06 $ &  0.09 & 3.1 & \nodata & \nodata \\
M05, $Z_{\odot}$, $\tau$                   &  $ -0.24\pm 0.07 $ & $ 0.27 \pm 0.04  $ & $ -0.17 \pm 0.06 $ &  0.05 & 3.7 & 0.15    & 0.36    \\
M05, $Z_{\odot}$, double burst (10\%)     &  $ -0.28\pm 0.07 $ & $ 0.29 \pm 0.03  $ & $ -0.21 \pm 0.07 $ &  0.07 & 2.2 & 6.1\%   & 1.74$^*$  \\
M05, $Z_{\odot}$, double burst (30\%)     &  $ -0.32\pm 0.07 $ & $ 0.27 \pm 0.03  $ & $ -0.25 \pm 0.06 $ &  0.05 & 2.3 & 23\%    & 1.41$^*$  \\
M05, $2.2Z_{\odot}$                        &  $ -0.08 \pm 0.08$ & $ 0.27 \pm 0.05  $ & $ -0.37 \pm 0.06 $ &  0.09 & 1.5 & \nodata & \nodata \\
\hline
\multicolumn{8}{c}{$z=1$: $b_{435}$ , $v_{606}$ , $i_{775}$ , $z_{850}$ , $J$}\\
\hline
                 
BC03, $Z_{\odot}$, $x=1.35$      &  $ -0.18 \pm 0.07 $ & $ 0.28 \pm 0.03  $ & $ -0.28 \pm 0.07 $ & 0.30 & 1.8 & \nodata & \nodata \\
BC03, $\tau$, $A_V$              &  $ -0.05 \pm 0.07 $ & $ 0.24 \pm 0.03  $ & $ -0.15 \pm 0.06 $ & 0.20 & 2.1 & 0.29    & 0.28 \\
BC03, double burst, $A_V$        &  $ -0.03 \pm 0.07 $ & $ 0.28 \pm 0.03  $ & $ -0.12 \pm 0.06 $ & 0.24 & 2.7 & 15\%    & 0.60$^*$ \\
M05, $Z_{\odot}$, $x=1.35$       &  $ -0.31 \pm 0.07 $ & $ 0.28 \pm 0.04  $ & $ -0.41 \pm 0.06 $ & 0.24 & 1.4 & \nodata & \nodata \\
M05, $\tau$, $A_V$               &  $ -0.23 \pm 0.07 $ & $ 0.25 \pm 0.03  $ & $ -0.33 \pm 0.06 $ & 0.22 & 1.7 & 0.13    & 0.22 \\
\enddata
\tablecomments{
SED fitting results for the high-$z$ sample.
The photometry used in the fits is indicated for the $z\sim 1$ sample.
For the $z=0$ sample similar rest-frame wavelength coverages are used (see text for the exact filter sets).
$\delta_1$ is the average $\log(M_{\rm{phot}}/M_{dyn})$, and $\sigma_1$ is the scatter therein.
$\Delta\equiv \delta_1-\delta_0$.
We use $\delta_0$ from single burst models to calculate $\Delta$, also in case
of more complicated SFHs and dust-extinction.
$x=0.35$ indicates models with a flat IMF (with slope 0.35 instead of 1.35 which is the slope of the Salpeter IMF).
$SFH$ indicates either the average time-scale $\tau$ of the exponentially declining star-formation rate,
or the average percentage of stellar mass involved in a secondary burst. This percentage
ranges between 0\% and the adopted burst strength, which indicates that not all galaxies
are necessarily fit better by a double burst model than by a single burst model.
$A_V$ is the extinction in the rest-frame $V$ band. 
Those values marked with a $*$ indicate the attenuation of the secondary burst, the 
primary burst is assumed to be dust-free.
\label{tab2}}
\end{deluxetable*}

\begin{figure*}
\begin{center}
\leavevmode
\hbox{%
\epsfxsize=18cm
\epsffile{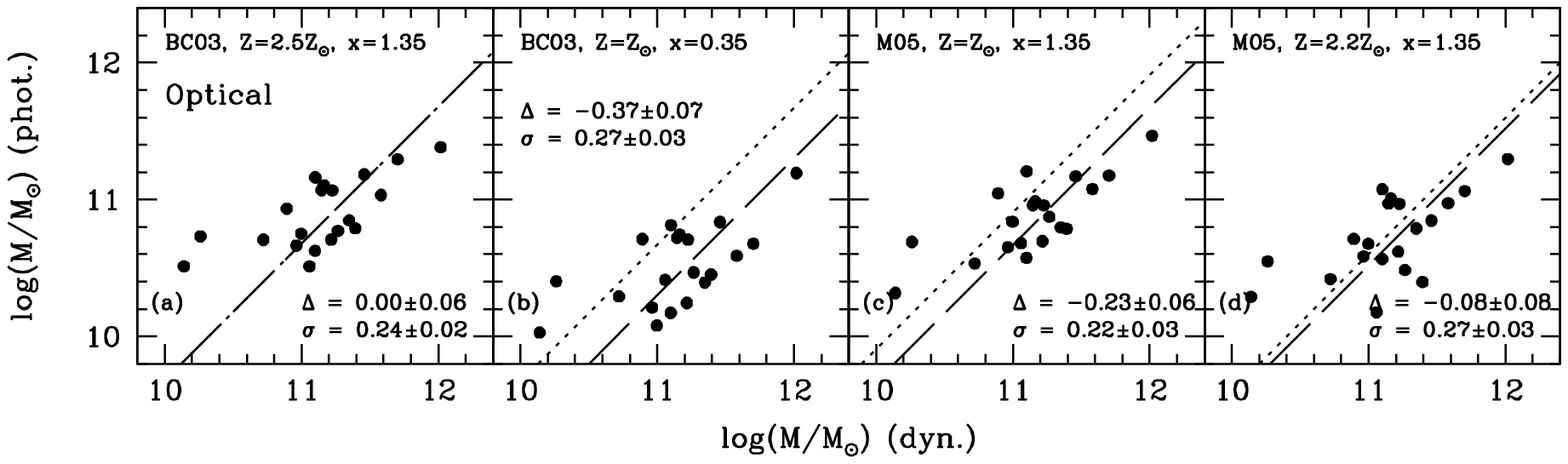}}
\figcaption{\small
Comparison between dynamical and photometric masses
of distant field early-type galaxies as in Figures \ref{MMstandardopt}, 
including $i_{775}$, $z_{850}$, and $J$ in the fits.
The photometric masses are obtained by using different models.
Panel (a) shows the results for the BC03 model
with super-solar metallicity ($2.5Z_{\odot}$) and a Salpeter IMF.
Panel (b) shows the results for the BC03 model
with solar metallicity and a top-heavy IMF 
(with $x=0.35$ instead of $x=1.35$ (Salpeter)).
Panel (c) shows the results from the M05 model with solar 
metallicity and a Salpeter IMF. 
Panel (d) shows the results from the M05 model with $2.2Z_{\odot}$
and a Salpeter IMF.
$\Delta$ is inconsistent with zero in panels
a and c, which indicates a discrepancy between the models
and the observed SEDs and $M/L$.
Also, we note that $\delta_i$ is smaller than zero in all cases
(the dashed and dotted lines lie below the line $M_{phot}=M_{dyn}$),
which indicates the presence of dark matter or a larger
number of low-mass stars than assumed in the Salpeter IMF.
\small
\label{MMmodels2}}
\end{center}
\end{figure*}

To quantify the difference between $M_{phot}$ and $M_{dyn}$,
we introduce the parameter $\delta_i = \log(M_{phot}/M_{dyn})$,
where the subscript $i=0$ or $i=1$ refers to the redshift of the sample.
$\delta_0 = -0.10\pm0.04$, which means that photometric masses are 
$20 \pm 8\%$ smaller than dynamical masses for the low-redshift sample.
For the high-$z$ sample we find $\delta_1=-0.20 \pm 0.07$.
The fact that $\delta_i<0$ may be explained by dark matter, or
an underestimate of the number of low-mass stars in these galaxies.
It is more relevant that $\delta_1$ and $\delta_0$ are consistent with 
each other:
the photometry does not allow us to constrain the low-mass
end of the IMF and the dark matter fraction.
Therefore, we normalize the low-$z$ photometric masses such that, on average, 
they are equal to the dynamical masses.
The $z\sim 1$ photometric mass estimates are changed by the same amount such
that the normalized offset between $M_{phot}$ and $M_{dyn}$ at $z\sim 1$
can be expressed as $\Delta = \delta_1 - \delta_0$.
Thus, for the above, we have $\Delta = -0.11 \pm 0.07 $, i.e.,
the normalized $z\sim 1$ photometric masses are consistent with the dynamical
masses (they are marginally smaller, by $22 \pm 14\%$).
The fact that $\Delta$ is consistent with zero means that the differences
between the SEDs of the low- and high-redshift samples are correctly
transformed into a difference in $M/L$ by this BC03 model.
\begin{figure}
\begin{center}
\leavevmode
\hbox{%
\epsfxsize=9cm
\epsffile{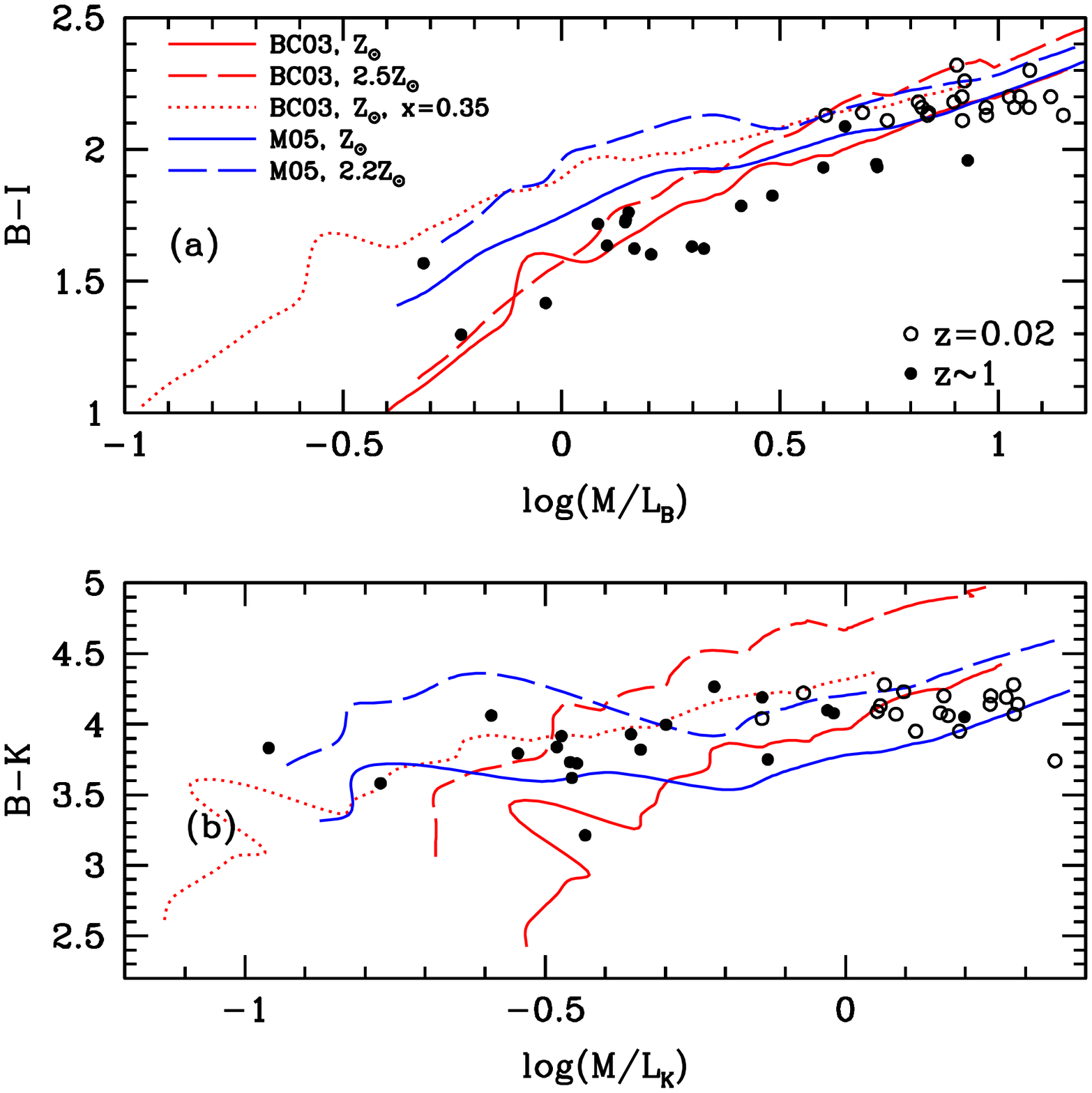}}
\figcaption{\small
The relation between color and $M/L$ for the models used to
infer photometric mass estimates,
compared with observed galaxy colors and dynamically 
obtained $M/L$.
Open circles represent the local early-type sample;
filled circles represent the $z\sim 1$ early-type sample.
Panel (a) shows $M/L_B$ and $B-I$.
The BC03 models with a Salpeter IMF agree better
with observations than the M05 models.
Panel (b) shows $M/L_K$ and $B-K$.
In this case, the M05 model agrees better with the data than the
BC03 models.
\small
\label{ML_col}}
\end{center}
\end{figure}
As can be seen, $M_{phot}$ and $M_{dyn}$ correlate well;
the scatter around the average $M_{phot}/M_{dyn}$
is a factor of 1.50 for the local sample and a factor of 1.78 for the 
distant sample
(these and all other modeling results for the high-$z$ sample are 
summarized in Table \ref{tab2}).

In order to investigate systematic effects,
we perform SED fits with different model parameters.
We show the results for the high-redshift galaxies in
Figure \ref{MMmodels2}.
If we fit a super-solar metallicity BC03 model (panel a) 
we obtain  $\delta_1=-0.32\pm0.05$ and $\Delta=0.00\pm 0.06$.
If we adopt a flat  IMF (panel b), we find $\delta_1=-0.70\pm  0.06$
and $\Delta=-0.37 \pm 0.07$.
Thus, if IMF and metallicity are unconstrained the average photometric
mass of the high-$z$ sample is uncertain by at least a factor of 5 in an absolute sense.
The differential $M/L$ from $z=1$ to $z=0$ is uncertain by a factor of 2.3.

Now we explore the M05 models, using the same ages, metallicities and
IMF as for the BC03 models.
In panel c of Figure \ref{MMmodels2} we show the $z\sim 1$ results
if we adopt the M05 model with solar metallicity and a Salpeter IMF 
(see also Table \ref{tab2}).
We find that $\Delta=-0.23\pm 0.06$, which is significantly smaller
than what was found for the BC03 model and also significantly smaller
than zero: 
the normalized high-$z$ photometric masses are a factor of $1.7\pm 0.2$ smaller
than the dynamical masses.
We find better agreement for the super-solar metallicity M05 model
(panel d of Figure \ref{MMmodels2}): $\Delta=-0.08\pm0.08$.
The disagreement between the BC03 and M05 models implies a systematic 
uncertainty of a factor of 1.3 in the normalized high-$z$ mass estimates.
This is much smaller than the uncertainty due to unconstrained metallicity
and IMF.
The scatter in $M_{phot}/M_{dyn}$ is similar to that obtained with the
BC03 models (a factor of 1.65-1.85).

To illustrate the results described above,
we show in Figure \ref{ML_col}a 
the relation between optical color ($B-I$) and $M/L$ for the models
and the observed galaxies.
The BC03 models with a Salpeter IMF reproduce the colors and $M/L$ rather well,
whereas the BC03 model with the flat IMF and the M05 models do not.
For the high-$z$ galaxies, the M05 model under-predicts $M/L$,
which explains that this model gives $\Delta<0$.

\begin{figure*}
\begin{center}
\leavevmode
\hbox{%
\epsfxsize=15cm
\epsffile{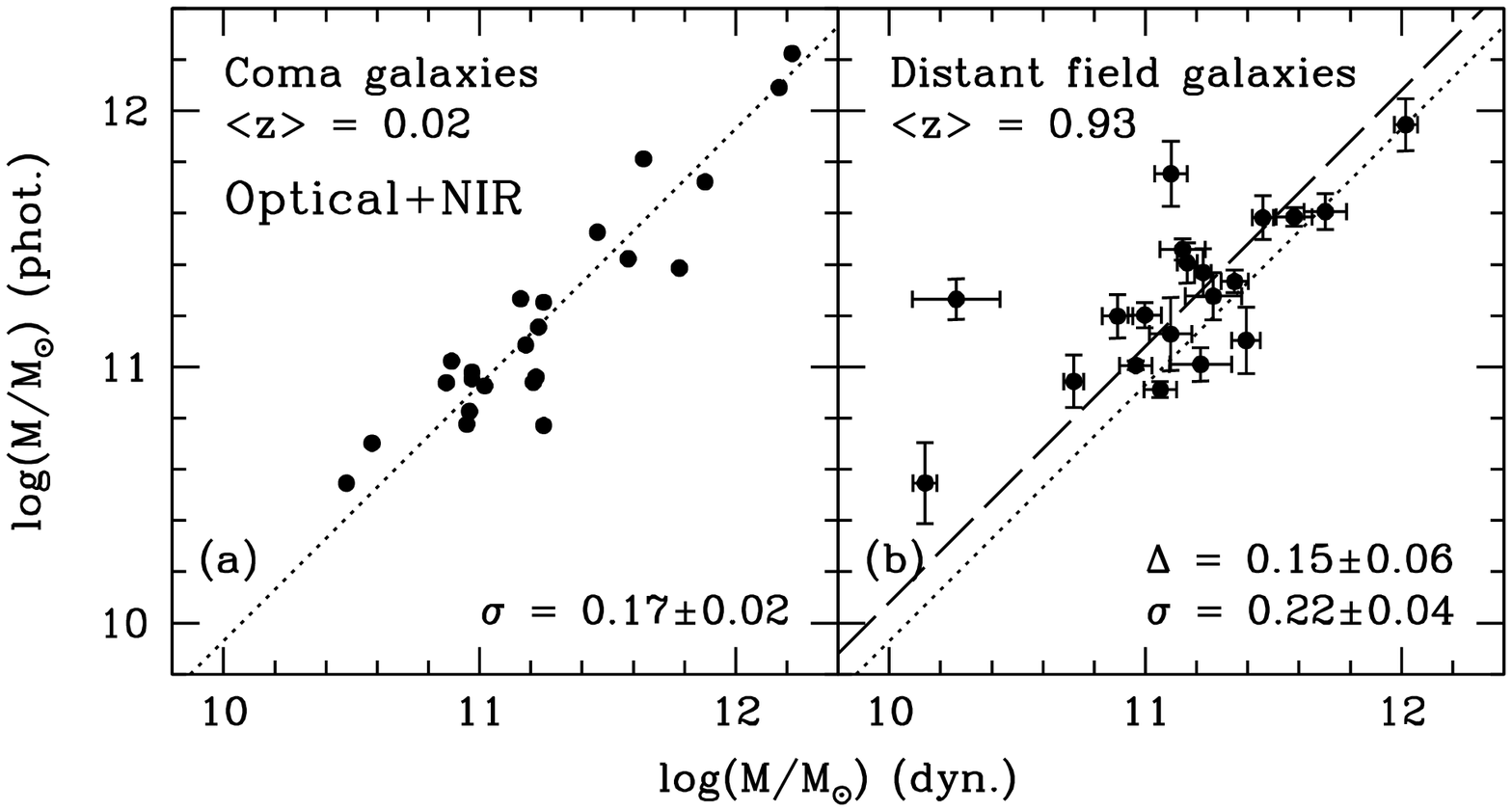}}
\figcaption{\small
A comparison between dynamical and photometric masses
as in Figure \ref{MMstandardopt},
but now including the rest-frame near-IR in the SED fits.
For the local Coma sample, the $B$-, $V$-, $I$-, and $K$-band
data-points are used.
The average photometric mass is not affected 
by the changed wavelength coverage:
the dotted line has not shifted significantly
with respect to Figure \ref{MMstandardopt}.
For the distant sample, $i_{775}$ through $4.5\mu$
are used to obtain photometric masses.
There is a systematic increase with respect 
to the masses obtained with the optical SEDs alone:
the dashed line has shifted with respect to 
Figure \ref{MMstandardopt}.
Now, $\Delta> 0$, which implies that stellar
masses at $z\sim 1$ are overestimated relative
to $z=0$.
\small
\label{MMstandard}}
\end{center}
\end{figure*}

\subsection{Rest-Frame Optical and Near-IR SED Fits}\label{optnir}

\begin{figure*}
\begin{center}
\leavevmode
\hbox{%
\epsfxsize=17cm
\epsffile{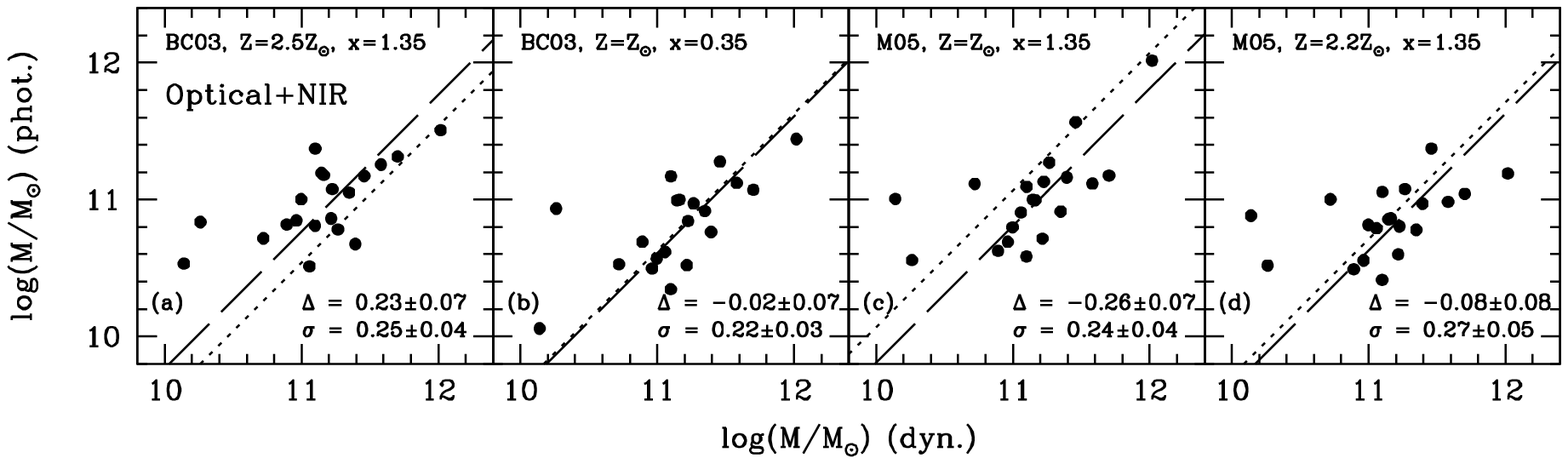}}
\figcaption{\small
A comparison between dynamical and photometric masses
as in Figure \ref{MMstandard} (fitting rest-frame
optical through near-IR SEDs). The same set of models as in 
Figure \ref{MMmodels2} are used.
For any BC03 model, $\Delta$ increases significantly
when the near-IR is included in the SED fits.
For the M05 model $\Delta$ does not significantly change.
\small
\label{MMmodels1}}
\end{center}
\end{figure*}

We now extend the SED fits to the rest-frame $K$-band. 
The local galaxies are fit with the $B$-, $V$-, $I$-, and $K$-band
photometry.
The distant galaxies are now fit with $i_{755}$ through $4.5\mu$.
Similar to Figure \ref{MMstandardopt}, we show in Figure \ref{MMstandard}
the results if we adopt the BC03 model with solar metallicity and a Salpeter
IMF.
We now find $\delta_1=0.08\pm 0.05$, i.e., the high-$z$ photometric
masses have increased by almost a factor of 2 with respect
to the photometric masses obtained from the optical SEDs alone.
This is not the case for the local sample ($\delta_0=-0.07\pm 0.03$),
which implies that $\Delta$ has increased significantly, 
to $\Delta=0.15\pm 0.06$.
Thus, the normalized high-$z$ stellar masses are now a factor of 1.4 larger
than the dynamical masses.
A similar increase in $\Delta$ is found for the other BC03 SSP models
(see panels a and b of Figure \ref{MMmodels1} and Table \ref{tab2}).
For the model with a flat IMF this means that $\Delta$ is consistent
with zero, but we have shown in the previous section 
that such a model provides a very poor fit to the rest-frame optical 
colors and $M/L$.

For the M05 SSP models (panels c and d of Figure \ref{MMmodels1}) 
the effect of including the near-IR is very different.
It is still true that the high-$z$ photometric 
masses increase somewhat, but that is also the case for the local sample.
The net effect is that $\Delta$ does not change significantly.
For the solar metallicity model we find $\Delta=-0.26\pm0.07$
and for the super-solar metallicity we find $\Delta=-0.08\pm0.08$.

The most important consequence of the above results is the large difference
between the masses inferred from the BC03 and M05 models 
with the same parameters.
Normalized masses at $z\sim 1$ obtained with a BC03 model are 2-2.5 
times larger than normalized masses obtained with a M05 model.
This implies that there is an intrinsic systematic uncertainty
in photometric mass estimates of at least a factor of 2.5.
We stress that our results are not caused by the different redshifts
of the samples: the same rest-frame wavelength is sampled
for the low- and high-redshift samples. The age difference between
the samples reveals the systematic problems.
In principle, tests such as described here do not require 
a range in redshift, but in the local universe there
is no suitable sample of relatively simple stellar systems with ages
of 1-2 Gyr and dynamically measured masses.

In Figure \ref{ML_col}b we show the relation between $B-K$ 
and $M/L_K$ for the models and the observed galaxies.
The M05 models reproduce the galaxy colors and $M/L$ 
better than the BC03 models.
Note that in the optical the reverse is the case 
(see Figure \ref{ML_col}a).
Qualitatively, we may understand that extending SED fits
to the near-IR does not constrain mass estimates much further:
the optical-to-near-IR color (e.g., $B-K$)
does not constrain $M/L$ very well.
Reversely, the M05 model, which predicts the lack of strong evolution
in $B-K$ cannot distinguish well between galaxies with low
and high $M/L$.
As a side-effect, the scatter in $M_{phot}/M_{dyn}$ is similar to that
found when the SED fits are restricted to the optical,
a factor of 1.65-1.85 for all models described above.

The relatively high masses inferred with the BC03 model
for the $z\sim 1$ galaxies 
are the result of the relatively high near-IR luminosities of the galaxies.
According to the model these imply old age and high $M/L$
(see Figures \ref{age_col} and \ref{ML_col}).
Another possible explanation for the high near-IR luminosities
is the presence of dust-obscured stellar populations that
do not contribute to the optical luminosity.
Generally speaking, models with more complex
star-formation histories might yield younger ages
and lower $M/L$.
We explore models with exponentially declining star-formation rates
and models with a dust-obscured,
secondary burst of star-formation that occurs
$2~\rm{Gyr}$ after the initial burst.

We adopt a fixed contribution to the stellar mass 
of the secondary burst and leave age and extinction
as free parameters.
Leaving the secondary burst strength as a free parameter
results in a very large scatter in $M_{phot}/M_{dyn}$
as the three free parameters are degenerate.
If we assume that a secondary burst 
accounts for 10\% of the total stellar mass,
we find $\Delta=0.07\pm 0.06$ (where $\delta_0$ is taken from
the results obtained with the SSP model as the effect of a secondary
burst does not affect the SEDs of the old local early-types). 
The galaxies responsible
for the decrease in $\Delta$ typically have a $0.1~\rm{Gyr}$
old secondary stellar population that is highly obscured ($A_V\sim 4$).
Similar results are obtained if the secondary burst
strength is increased to 30\% ($\Delta=0.01\pm0.07$).
Models with exponentially declining star-formation rates
also improve the results for the BC03 model if 
we allow for the presence of dust ($\Delta=0.03\pm 0.06$).
Hence, models with a more complex SFH can successfully
reproduce the SEDs of $z\sim 1$ early-type galaxies
that yield photometric masses that are consistent
with their dynamical masses.
Also, the quality of the fits is better for the
models with more complicated star-formation histories
than for the SSP model (see Table \ref{tab2}).
We note that these type of models do not improve the results
for the super-solar metallicity BC03 model ($\Delta>0.2$, see Table \ref{tab2}).
The M05 model fits do not change significantly
if a model with a complex star-formation history is used
($\Delta<-0.2$, see Table \ref{tab2}).

\subsection{Rest-Frame UV and Optical SED Fits: Constraints on the Star-Formation History}\label{uv}
Finally, we explore the effect of extending the analysis to 
the rest-frame UV. The UV is strongly affected by (small traces of) 
star-formation, and is expected to cause less accurate mass estimates. 
On the other hand, this sensitivity to young stars may allow to constrain 
the SFH and dust content, which we below adopt as free parameters.

\begin{figure*}
\begin{center}
\leavevmode
\hbox{%
\epsfxsize=17cm
\epsffile{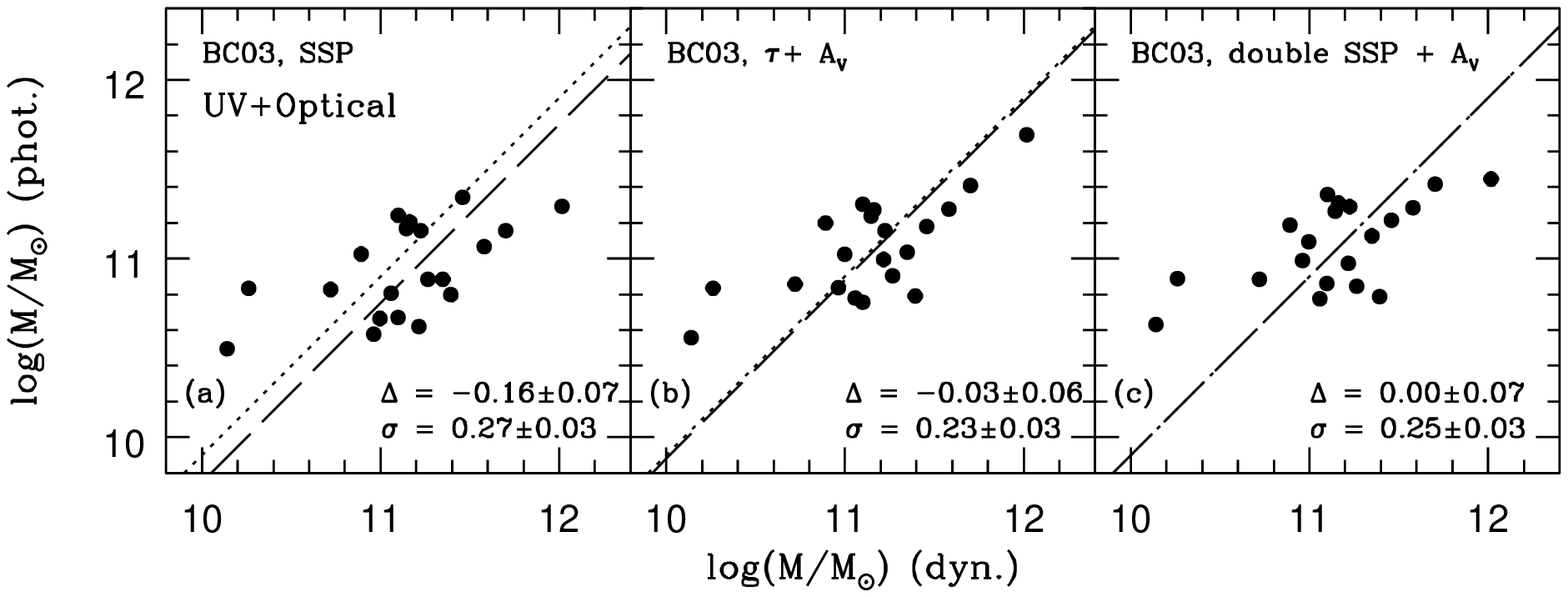}}
\figcaption{\small
Comparison between dynamical and photometric masses (using the BC03 model)
of distant field early-type galaxies as in Figure \ref{MMstandardopt}.
Here we modeled the rest-frame UV+optical SEDs:
all photometric data points between $b_{435}$ 
and $J$ are used in the fits.
In panel (a) we assume, as a starting point, a dust-free SSP
with solar metallicity and a Salpeter IMF.
In panel (b), $\tau$ and $A_V$ are free parameters.
In panel (c), the strength of a secondary burst that occurs
$2~\rm{Gyr}$ after the initial burst and $A_V$ 
(only applied to the secondary burst) are free parameters.
See Table \ref{tab2} for the inferred ages, and average values of 
$\tau$ and $A_V$.
\small
\label{MMUVBC}}
\end{center}
\end{figure*}

As a baseline, we adopt the BC03 and M05 SSP models with solar metallicity 
and a Salpeter IMF. We now fit the SEDs using the $b_{435}$, $v_{606}$, $i_{775}$, 
$z_{850}$, and $J$ photometry. Because of the problems detailed in the 
previous section, we omit the rest-frame near-IR. The results are shown in the 
left-hand panels of Figure \ref{MMUVBC} (BC03) and Figure \ref{MMUVM} 
(M05). The scatter in $M_{phot}/M_{dyn}$ is a factor of $\sim 1.9$,
similar to the scatter obtained without the $b_{435}$ and $v_{606}$ photometry.
The quality of the fits has decreased significantly: for the BC03 model, the 
average $\rm{RMS}$ is now 0.30 mag (it is 0.03 mag without the two 
shortest-wavelength filters); for the M05 model this is 0.24 mag, instead of 
0.05 mag. This indicates that, at least for the UV part of the SEDs, an SSP 
model provides poor fits.

Now we allow for an exponentially declining star-formation rate with 
timescale $\tau$, instead of assuming a single burst. 
Since star-formation is usually related to extinction by dust, 
we consequently adopt $A_V$ as a free parameter as well. 
The results are shown in the middle and panel of Figure \ref{MMUVBC} 
(BC03) and the right-hand panel of Figure \ref{MMUVM} (M05). 
The scatter in $M_{phot}/M_{dyn}$ decreases marginally and is 
now comparable (a factor of $\sim 1.7$) 
to what we found when we excluded the UV 
and assumed a dust-free SSP. 
Also, the fits are somewhat better,
as the average $\rm{RMS}$ is now $\sim 0.2~\rm{mag}$ for both models.
The fact that the $\rm{RMS}$ remains much higher 
than for the optical SED fits is
most likely caused by the large errors in the $b_{435}$-band photometry. 

In Table \ref{tab2} we show the average values of age, $\tau$, and $A_V$.
The inferred values for the extinction are quite large: $A_V\sim 0.25$ for both 
models. Age and $\tau$ are $\sim 2.1$ and $\sim 0.29~Gyr$ for the BC03 model 
with dust and $\sim1.7$ and $\sim 0.13~Gyr$ for the M05 model with dust. This 
implies an average SFR of $\sim 1~M_{\odot}~yr^{-1}$. Assuming that this 
continues to decline 
exponentially between the epoch of observation and the present, the 
average stellar mass increases by no more than $\sim 0.2\%$ after $z\sim 1$.
The galaxy with the highest SFR would increase its stellar mass by $8.0\%$.
The models imply that residual star-formation is not very relevant for the 
galaxies in our sample, and, barring interactions or mergers, have assembled
most of their stellar mass by $z\sim 1$.

Finally, we test a BC03 model in which a dusty, secondary burst of 
star-formation occurs $2~\rm{Gyr}$ after the dust-free initial burst. 
This is the same model as used in the previous section, but now we adopt
the strength of the secondary burst as a free parameter as well, along
with age and $A_V$:
we allow the secondary burst to
vary in strength between 0\% to 30\% of the final stellar mass.
The results are shown in the right-hand panel of Figure 
\ref{MMUVBC}. The scatter in 
$M_{phot}/M_{dyn}$ is a factor of 1.9, which is similar to what was
found when modeling the rest-frame optical SEDs with a dust-free SSP model.

To investigate the applicability of the double burst model, it is more 
interesting to look at individual galaxies. The fits of 11 out of 20 
galaxies clearly improve by allowing a dusty, secondary burst. With a single burst, 
these seven galaxies are assigned ages of $1.2~\rm{Gyr}$. 
These galaxies are fit significantly better by a $\sim 0.3~\rm{Gyr}$
old secondary burst involving, on average, $12\%$ of the final stellar mass
and an extinction of $A_V\sim 0.65$.
This young population is superimposed on a 2 Gyr older, dust-free population.

\section{Summary and Discussion}\label{discussion}
In this paper we have compared photometric and dynamical
masses of early-type galaxies at both high ($z\sim 1$) and low redshifts.
The uncertainties were analyzed quantitatively. 
We start this discussion with summarizing our main conclusions.

1) 
After allowing for a systematic offset, 
the scatter in $M_{phot}/M_{dyn}$, which quantifies the random uncertainty 
in photometric mass estimates of individual
early-type galaxies at $z\sim 1$, is a factor of $\sim 1.75$ ($\sigma=0.25$). 
We find no significantly different random uncertainties for different models
and different wavelength ranges used in the fits, 
i.e., including IRAC data in the SED fits does not provide
significantly more accurate mass estimates.

2) For SSP models from BC03 and rest-frame optical SED fits we find that the 
photometric masses of the $z\sim 1$ galaxies are consistent
with the dynamical masses after normalization with respect to
the $z=0$ galaxies. 
This normalization is the difference between 
the photometric and dynamical masses of the low-$z$ comparison sample.
When the SED fits are extended to the rest-frame near-IR
the normalized $z\sim 1$ mass estimates increase by a factor of 2:
the $z\sim 1$ photometric masses are are larger than the dynamical masses.
For the SSP model with a top-heavy IMF ($x=0.35$ instead of $x=1.35$)
there is no such discrepancy, but in that case
optical SED fits underestimate the $z\sim 1$ masses by a factor of 2.3.

If we allow for more complex star-formation histories and dust-extinction, 
the $z\sim 1$ mass estimates decrease, 
such that the discrepancy seen for the BC03 models with a Salpeter IMF
and solar metallicity disappears.
For example, a model with a secondary, obscured burst of star formation
provides good results and increases the quality of the fits.
\null
\vbox{
\begin{center}
\leavevmode
\hbox{%
\epsfxsize=8.5cm
\epsffile{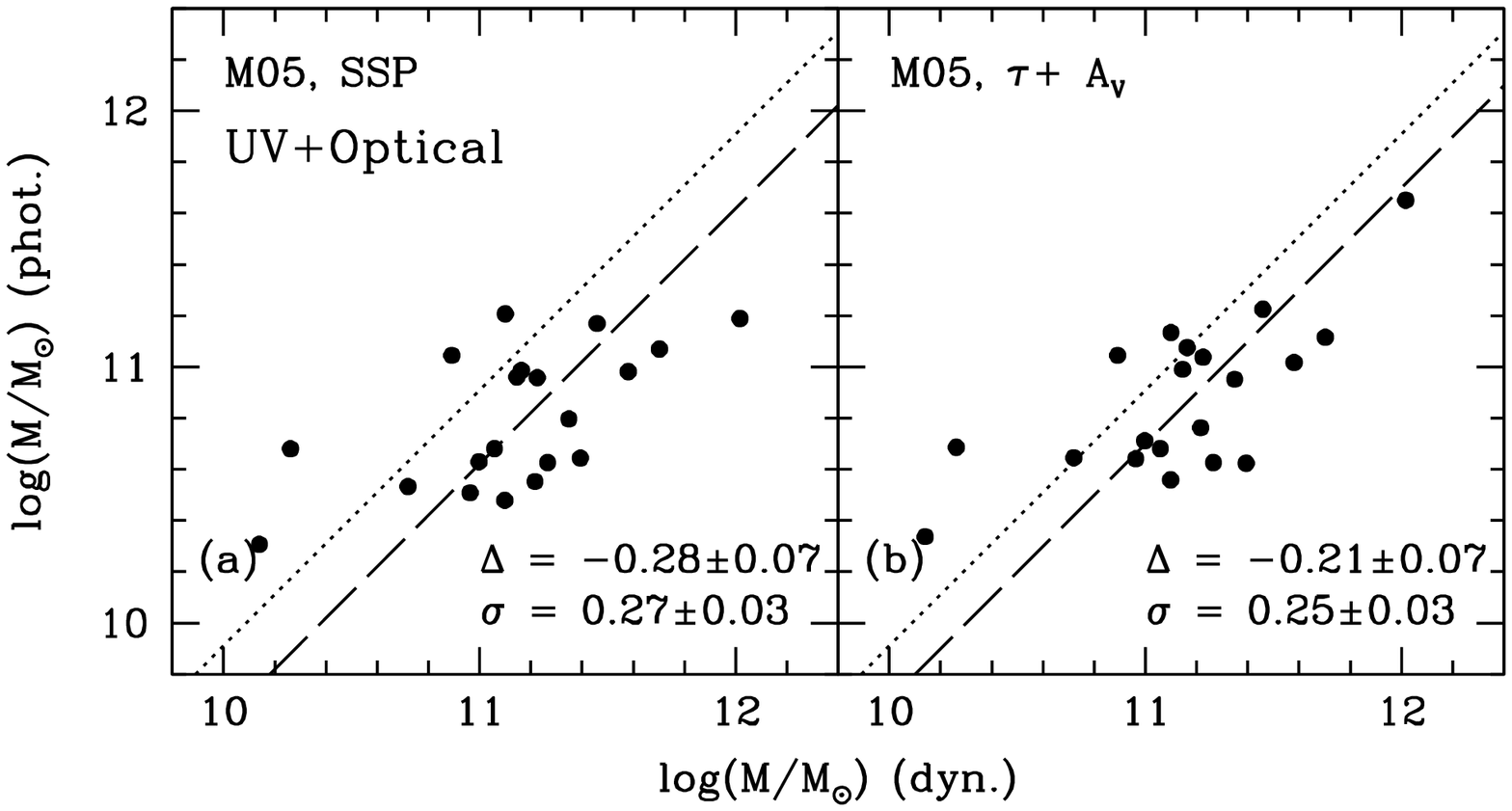}}
\figcaption{\small
Comparison between dynamical and photometric masses (using the M05 model)
of distant field early-type galaxies as in Figure \ref{MMstandardopt},
modeling the rest-frame UV+optical SEDs: all photometric data points 
between $b_{435}$ and $J$ are used in the fits.
In panel (a) we assume, as a starting point, a dust-free SSP.
In panel (b), $\tau$ and $A_V$ are free parameters.
See Table \ref{tab2} for the best-fitting model properties.
\small
\label{MMUVM}}
\end{center}}
If correct, these results imply the presence of a significant 
population (at least 10\% in mass) of young stars ($\sim 0.1~\rm{Gyr}$)
that are highly obscured ($A_V\sim 4$) in roughly half of
the $z\sim 1$ population. 
It remains to be seen whether this is the case.
Mid-IR observations will constrain this scenario.

3)
We find very different results when we use the M05 model with solar 
metallicity.
That model produces normalized mass estimates from the optical SEDs
of the $z\sim 1$ galaxies that are too low by a factor of 1.7.
The most striking difference with the BC03 model
is revealed when the SED fits are extended to the near-IR:
the normalized mass estimates at $z\sim 1$ do not change significantly,
which implies that the normalized high-$z$ mass estimates remain too low.
More complex models do not change these results significantly.
Most importantly, the differences between the results
obtained with the BC03 and M05 SSP models with identical
model parameters imply a systematic uncertainty
of a factor of 2.5 in photometric mass estimates obtained from near-IR 
SED fits.
Hence, extending SED fits from the optical to the near-IR
does not provide better mass estimates of high-$z$
early-type galaxies, because of uncertainties
intrinsic to the stellar population models.

4) Adopting the SFH and dust-extinction as free parameters, and
extending the SED fits to the rest-frame UV, do not 
increase the quality of the SED fits, nor
decrease the random uncertainty in the photometric masses.
For our sample, fitting the rest-frame optical SEDs is optimal to constrain 
the masses.
Fits of UV+optical SEDs of several individual galaxies, however, 
improve significantly by adopting the SFH as a free parameter.

The above results demonstrate that both models
have problems in either the optical (M05) or the near-IR (BC03,
barring significant, obscured stellar populations).
This conclusion relies on two assumptions.
First, we assume that all local and distant galaxies have similar dynamical 
structures.
In other words, we assume that invoking $5r_{\rm{eff}}\sigma_c^2 /G$
as $M_{dyn}$ for both low- and high-redshift early-types 
does not introduce systematic effects.
This assumption appears to be reasonable, as FP studies
have shown that all nearby ellipticals and S0s to follow the same relation 
\citep[e.g.][]{jorgensen96}, and \citet{treu04} showed
that high redshift ellipticals have the same structure as low redshift
ellipticals. 
Second, we assume that local and distant galaxies can be fit with 
models with the same metallicity and IMF.
By that we ignore, for example, the possibility that the galaxies
in our $z\sim 1$ sample evolve strongly in metallicity. 
However, in order to remove the inconsistency found for the BC03 model,
the distant galaxies would have to be more metal-rich than the local
galaxies.
This seems unlikely. 
On the contrary, \citet{jorgensen05} show evidence for metal enrichment
between $z\sim 0.8$ and the present.
We also note that mergers between early-type galaxies, 
which are likely very important in 
shaping the mass-function of the early-type galaxy population
\citep[e.g.,][]{vandokkum05},
do not affect our conclusions much because the stellar populations 
themselves do not change.

The differences between the models are not restricted to the range of ages
of the galaxies described in this paper ($\gtrsim 1~\rm{Gyr}$). 
On the contrary, the largest differences are found for even younger ages 
(i.e., low $M/L$, see Figures \ref{age_col} and \ref{ML_col}). 
This implies that mass estimates of young or star-forming
galaxies may also be systematically uncertain. 
This is in particular relevant to
studies at $z>2$, where SED fits are the only available method for estimating
galaxy masses 
\citep[e.g.,][]{forster04,daddi04,shapley05,labbe05,mobasher05}.
This work suggests that more than one type of model should be used
when performing SED fits, 
as the difference will give an indication of the systematic uncertainty. 

It is essential that the models are improved, hopefully
leading to consistent predictions of the SEDs, given the
IMF, star formation history and metallicity.
Therefore, a better understanding of the near-IR properties of stellar populations is required.
Our work provides a new tool to verify the predictions of the models,
a method that can be exploited more thoroughly by
obtaining a large sample of early-type galaxies with a wide range in redshift,
accurate dynamical masses and abundance measurements from spectroscopy,
star-formation activity from UV and mid-IR photometry, and mass estimates 
derived from multi-wavelength imaging, including the rest-frame near-IR.
Only when the model discrepancies are resolved can the full potential of IRAC
for measuring stellar masses of distant galaxies be realized.


\begin{thebibliography}{41}
\expandafter\ifx\csname natexlab\endcsname\relax\def\natexlab#1{#1}\fi
\bibitem[{{Bell} \& {de Jong}(2001)}]{bell01}
{Bell}, E.~F., \& {de Jong}, R.~S. 2001, \apj, 550, 212

\bibitem[{{Bell et al.}(2004)}]{bell04b}
{Bell, E.~F. et al.} 2004, \apj, 608, 752

\bibitem[{{Bruzual} \& {Charlot}(2003)}]{bruzual03}
{Bruzual}, G., \& {Charlot}, S. 2003, \mnras, 344, 1000

\bibitem[{{Cappellari et al.}(2005)}]{cappellari05}
{Cappellari, M. et al.} 2005, \mnras, 366, 1126

\bibitem[{{Cassisi et al.}(2000)}]{cassisi00}
{Cassisi}, S., {Castellani}, V., {Ciarcielluti}, P., {Piotto}, G., \& {Zocalli}, M.
2000, \mnras, 315, 679

\bibitem[{{Daddi et al.}(2004)}]{daddi04}
{Daddi}, E. et al. 2004, \apj, 600, L127

\bibitem[{{Dickinson et al.}(2003)}]{dickinson03}
{Dickinson}, M., {Papovich}, C., {Ferguson}, H.C., {Bud\'avari}, T. 
2003, \apj, 587, 25

\bibitem[{{Djorgovski} \& {Davis}(1987)}]{djorgovskidavis87}
{Djorgovski}, S., \& {Davis}, M. 1987, \apj, 313, 59

\bibitem[{{Dressler} {et~al.}(1987){Dressler}, {Lynden-Bell}, {Burstein},
  {Davies}, {Faber}, {Terlevich}, \& {Wegner}}]{dressler87}
{Dressler}, A., {Lynden-Bell}, D., {Burstein}, D., {Davies}, R.~L., {Faber},
  S.~M., {Terlevich}, R., \& {Wegner}, G. 1987, \apj, 313, 42

\bibitem[{{Drory} {et~al.}(2004{\natexlab{a}}){Drory}, {Bender}, {Feulner},
  {Hopp}, {Maraston}, {Snigula}, \& {Hill}}]{drory04}
{Drory}, N., {Bender}, R., {Feulner}, G., {Hopp}, U., {Maraston}, C.,
  {Snigula}, J., \& {Hill}, G.~J. 2004{\natexlab{a}}, \apj, 608, 742

\bibitem[{{Drory} {et~al.}(2004{\natexlab{b}}){Drory}, {Bender}, \&
  {Hopp}}]{drory04b}
{Drory}, N., {Bender}, R., \& {Hopp}, U. 2004{\natexlab{b}}, \apjl, 616, L103

\bibitem[{{Faber} {et~al.}(1989){Faber}, {Wegner}, {Burstein}, {Davies},
  {Dressler}, {Lynden-Bell}, \& {Terlevich}}]{faber89}
{Faber}, S.~M., {Wegner}, G., {Burstein}, D., {Davies}, R.~L., {Dressler}, A.,
  {Lynden-Bell}, D., \& {Terlevich}, R.~J. 1989, \apjs, 69, 763

\bibitem[{{Faber et al.}(2005)}]{faber05}
{Faber, S.~M. et al.} 2005, \apj, submitted, astro-ph/0506044

\bibitem[{{Fazio} {et~al.}(2004){Fazio}, {Hora}, {Allen}, {Ashby}, {Barmby},
  {Deutsch}, {Huang}, {Kleiner}, {Marengo}, {Megeath}, {Melnick}, {Pahre},
  {Patten}, {Polizotti}, {Smith}, {Taylor}, {Wang}, {Willner}, {Hoffmann},
  {Pipher}, {Forrest}, {McMurty}, {McCreight}, {McKelvey}, {McMurray}, {Koch},
  {Moseley}, {Arendt}, {Mentzell}, {Marx}, {Losch}, {Mayman}, {Eichhorn},
  {Krebs}, {Jhabvala}, {Gezari}, {Fixsen}, {Flores}, {Shakoorzadeh}, {Jungo},
  {Hakun}, {Workman}, {Karpati}, {Kichak}, {Whitley}, {Mann}, {Tollestrup},
  {Eisenhardt}, {Stern}, {Gorjian}, {Bhattacharya}, {Carey}, {Nelson},
  {Glaccum}, {Lacy}, {Lowrance}, {Laine}, {Reach}, {Stauffer}, {Surace},
  {Wilson}, {Wright}, {Hoffman}, {Domingo}, \& {Cohen}}]{fazio04}
{Fazio}, G.~G. et al. 2004, \apjs, 154, 10

\bibitem[{{F{\" o}rster Schreiber} {et~al.}(2004){F{\" o}rster Schreiber}, {van
  Dokkum}, {Franx}, {Labb{\' e}}, {Rudnick}, {Daddi}, {Illingworth}, {Kriek},
  {Moorwood}, {Rix}, {R{\" o}ttgering}, {Trujillo}, {van der Werf}, {van
  Starkenburg}, \& {Wuyts}}]{forster04}
{F{\" o}rster Schreiber}, N.~M. et al. 2004, \apj, 616, 40

\bibitem[{{Franx}(1993)}]{franx93}
{Franx}, M. 1993, \pasp, 105, 1058

\bibitem[{{Gerhard} {et~al.}(2001){Gerhard}, {Kronawitter}, {Saglia}, \&
  {Bender}}]{gerhard01}
{Gerhard}, O., {Kronawitter}, A., {Saglia}, R.~P., \& {Bender}, R. 2001, \aj,
  121, 1936

\bibitem[{{Giavalisco} {et~al.}(2004){Giavalisco}, {Ferguson}, {Koekemoer},
  {Dickinson}, {Alexander}, {Bauer}, {Bergeron}, {Biagetti}, {Brandt},
  {Casertano}, {Cesarsky}, {Chatzichristou}, {Conselice}, {Cristiani}, {Da
  Costa}, {Dahlen}, {de Mello}, {Eisenhardt}, {Erben}, {Fall}, {Fassnacht},
  {Fosbury}, {Fruchter}, {Gardner}, {Grogin}, {Hook}, {Hornschemeier}, {Idzi},
  {Jogee}, {Kretchmer}, {Laidler}, {Lee}, {Livio}, {Lucas}, {Madau},
  {Mobasher}, {Moustakas}, {Nonino}, {Padovani}, {Papovich}, {Park},
  {Ravindranath}, {Renzini}, {Richardson}, {Riess}, {Rosati}, {Schirmer},
  {Schreier}, {Somerville}, {Spinrad}, {Stern}, {Stiavelli}, {Strolger},
  {Urry}, {Vandame}, {Williams}, \& {Wolf}}]{giavalisco04}
{Giavalisco}, M. et al. 2004, \apjl, 600, L93

\bibitem[{{Girardi} {et~al.}(2000)}]{girardi00}
{Girardi}, L., {Bressan}, A., {Bertelli}, \& {Chiosi}, C. 2000, \aaps, 141, 371

\bibitem[{{J\o rgensen} {et~al.}(1996){J\o rgensen}, {Franx}, \&
  {Kjaergaard}}]{jorgensen96}
{J\o rgensen}, I., {Franx}, M., \& {Kjaergaard}, P. 1996, \mnras, 280, 167

\bibitem[{{J{\o}rgensen} {et~al.}(2005){J{\o}rgensen}, {Bergmann}, {Davies},
  {Barr}, {Takamiya}, \& {Crampton}}]{jorgensen05}
{J{\o}rgensen}, I., {Bergmann}, M., {Davies}, R., {Barr}, J., {Takamiya}, M.,
  \& {Crampton}, D. 2005, \aj, 129, 1249

\bibitem[{{Kochanek}(1994)}]{kochanek94}
{Kochanek}, C.~S. 1994, \apj, 436, 56

\bibitem[{{Kurucz}(1992)}]{kurucz92}
{Kurucz}, R.~L. 1992, in IAU Symp. 149: The Stellar Populations of Galaxies,
  225--+

\bibitem[{{Labb{\' e}} {et~al.}(2005){Labb{\' e}}, {Huang}, {Franx}, {Rudnick},
  {Barmby}, {Daddi}, {van Dokkum}, {Fazio}, {Schreiber}, {Moorwood}, {Rix},
  {R{\" o}ttgering}, {Trujillo}, \& {van der Werf}}]{labbe05}
{Labb{\' e}}, I. et al. 2005, \apjl, 624, L81

\bibitem[{{Maraston}(2005)}]{maraston05}
{Maraston}, C. 2005, \mnras, 362, 799

\bibitem[{{Mobasher} {et~al}(2005){{Mobasher}, B. and {Dickinson}, M. and {Ferguson}, H.~C. and 
	{Giavalisco}, M. and {Wiklind}, T. and {Stark}, D. and {Ellis}, R.~S. and 
	{Fall}, S.~M. and {Grogin}, N.~A. and {Moustakas}, L.~A. and 
	{Panagia}, N. and {Sosey}, M. and {Stiavelli}, M. and {Bergeron}, E. and 
	{Casertano}, S. and {Ingraham}, P. and {Koekemoer}, A. and {Labb{\'e}}, I. and 
	{Livio}, M. and {Rodgers}, B. and {Scarlata}, C. and {Vernet}, J. and 
	{Renzini}, A. and {Rosati}, P. and {Kuntschner}, H. and {K{\"u}mmel}, M. and 
	{Walsh}, J.~R. and {Chary}, R. and {Eisenhardt}, P. and {Pirzkal}, N. and 
	{Stern}, D.}}]{mobasher05}
{Mobasher}, B. et al. 2005, 635, 832

\bibitem[{{Pahre} {et~al.}(1998){Pahre}, {Djorgovski}, \& {de
  Carvalho}}]{pahre98}
{Pahre}, M.~A., {Djorgovski}, S.~G., \& {de Carvalho}, R.~R. 1998, \aj, 116,
  1591

\bibitem[{{Papovich} {et~al.}(2003){Papovich}, {Giavalisco}, {Dickinson},
  {Conselice}, \& {Ferguson}}]{papovich03}
{Papovich}, C., {Giavalisco}, M., {Dickinson}, M., {Conselice}, C.~J., \&
  {Ferguson}, H.~C. 2003, \apj, 598, 827

\bibitem[{{Rudnick et al.}(2003)}]{rudnick03}
{Rudnick, G. et al.} 2003, \apj, 599, 847

\bibitem[{{Rudnick et al.}(2006)}]{rudnick06}
{Rudnick, G. et al.} 2006, \apj, in press, astro-ph/0606536

\bibitem[{{Scodeggio} {et~al.}(1998){Scodeggio}, {Giovanelli}, \&
  {Haynes}}]{scodeggio98}
{Scodeggio}, M., {Giovanelli}, R., \& {Haynes}, M.~P. 1998, \aj, 116, 2728

\bibitem[{{Shapley} {et~al.}(2003){Shapley}, {Steidel}, {Pettini}, \&
  {Adelberger}}]{shapley03}
{Shapley}, A.~E., {Steidel}, C.~C., {Pettini}, M., \& {Adelberger}, K.~L. 2003,
  \apj, 588, 65

\bibitem[{{Shapley et al.}(2005)}]{shapley05}
{Shapley, A.~E. et al.} 2005, \apj, 626, 698

\bibitem[{{Trager et al.}(2000)}]{trager00}
{Trager}, S.~C. and {Faber}, S.~M. and {Worthey}, G. and {Gonz{\'a}lez}, J.~J. 2000, \apj, 120, 165

\bibitem[{{Treu} {et~al.}(2005){Treu}, {Ellis}, {Liao}, {van Dokkum}, {Tozzi},
  {Coil}, {Newman}, {Cooper}, \& {Davis}}]{treu05b}
{Treu}, T. et al. 2005, \apj, 633, 174

\bibitem[{{Treu} \& {Koopmans}(2004)}]{treu04}
{Treu}, T., \& {Koopmans}, L.~V.~E. 2004, \apj, 611, 739

\bibitem[{{Tully} \& {Fisher}(1977)}]{tully77}
{Tully}, R.~B., \& {Fisher}, J.~R. 1977, \aap, 54, 661

\bibitem[{{van der Wel} {et~al.}(2006){van der Wel}, {Franx},
  {van Dokkum}, {Huang}, {Rix}, \& {Illingworth}}]{vanderwel06}
{van der Wel}, A., {Franx}, M., {van Dokkum}, P.~G., {Huang}, J., {Rix}, H.-W.,
  \& {Illingworth}, G. 2006, \apj, 636, L21

\bibitem[{{van der Wel} {et~al.}(2005){van der Wel}, {Franx},
  {van Dokkum}, {Rix}, {Illingworth}, \& {Rosati}}]{vanderwel05}
{van der Wel}, A., {Franx}, M., {van Dokkum}, P.~G., {Rix}, H.-W.,
  {Illingworth}, G., \& {Rosati}, P. 2005, \apj, 631, 145

\bibitem[{{van Dokkum}(2005)}]{vandokkum05}
{van Dokkum}, P.~G. 2005, \aj, 130, 2647

\bibitem[{{van Dokkum} \& {Stanford}(2003)}]{vandokkumstanford03}
{van Dokkum}, P.~G., \& {Stanford}, S.~A. 2003, \apj, 585, 78

\end{thebibliography}
\end{document}